\documentclass[aps,pre,preprint,floatfix]{revtex4}

\usepackage[normalem]{ulem}
\usepackage{latexsym}
\usepackage[usenames]{color}
\usepackage{amsthm}
\usepackage{hyperref}
\usepackage{braket}

\usepackage{amsmath}    
\usepackage{amsfonts}  
\usepackage{amssymb}
\usepackage{graphicx}
\usepackage{slashed}
\usepackage{fouridx}
\usepackage{tikz}
\usetikzlibrary{calc}

\theoremstyle{plain} 
\theoremstyle{plain} 
\theoremstyle{plain}  
\theoremstyle{plain} 
\theoremstyle{plain}  
\theoremstyle{remark} 
\theoremstyle{plain} 
\theoremstyle{remark}

\newcommand\CROSS[1]{%
  \hbox{%
    \vbox{
      \hrule
      \kern2.5pt
      \hbox{$#1$\,\strut}
    }%
  \vrule
  }\mskip\thickmuskip
}

\tikzset{
solid node/.style={circle,draw,inner sep=1.5,fill=black},
hollow node/.style={circle,draw,inner sep=1.5}
}

\newlength{\arrowsize}  
\pgfarrowsdeclare{biggertip}{biggertip}{  
  \setlength{\arrowsize}{0.5pt}  
  \addtolength{\arrowsize}{.5\pgflinewidth}  
  \pgfarrowsrightextend{0}  
  \pgfarrowsleftextend{-5\arrowsize}  
}{  
  \setlength{\arrowsize}{0.5pt}  
  \addtolength{\arrowsize}{.5\pgflinewidth}  
  \pgfpathmoveto{\pgfpoint{-5\arrowsize}{4\arrowsize}}  
  \pgfpathlineto{\pgfpointorigin}  
  \pgfpathlineto{\pgfpoint{-5\arrowsize}{-4\arrowsize}}  
  \pgfusepathqstroke  
}

\begin{document}
\begin{center}
\Large{\textbf{Unifying vectors and matrices of different dimensions \\ through nonlinear embeddings}}\\ 
~\\

\large{Vladimir Garc\'{\i}a-Morales}\\


Departament de F\'{\i}sica de la Terra i Termodin\`amica\\ Universitat de Val\`encia \\ E-46100 Burjassot, Spain
\\ garmovla@uv.es
\end{center}
\small{}
\noindent  \normalsize{}
~\\ Complex systems may morph between structures with different dimensionality and degrees of freedom. As a tool for their modelling, nonlinear embeddings are introduced that encompass objects with different dimensionality as a continuous parameter $\kappa \in \mathbb{R}$ is being varied, thus allowing the unification of vectors, matrices and tensors in single mathematical structures. This technique is applied to construct warped models in the passage from supergravity in 10 or 11-dimensional spacetimes to 4-dimensional ones. We also show how nonlinear embeddings can be used to connect cellular automata (CAs) to coupled map lattices (CMLs) and to nonlinear partial differential equations, deriving a class of nonlinear diffusion equations. Finally, by means of nonlinear embeddings we introduce CA connections, a class of CMLs that connect any two arbitrary CAs in the limits $\kappa \to 0$ and $\kappa \to \infty$ of the embedding.

~\\

\pagebreak

\section{Introduction}

Scalars, vectors and matrices are mathematical objects with different numbers of dimensions (degrees of freedom) which generally store different amounts of information. The question whether these different mathematical structures can be encompassed by more general ones may be interesting in the modelling of complex systems (since these are often able to morph as continuous parameters are being varied \cite{Oliver}).  To that aim, we construct in this article mathematical structures that are able to behave as scalars, vectors or matrices when a parameter $\kappa \in \mathbb{R}$ is being continuously tuned from $0$ to $\infty$. These structures are specific instances of nonlinear $\mathcal{B}_{\kappa}$-embeddings that have been very recently introduced and applied to the problem of finding all roots of a complex polynomial \cite{homotopon}.

We can apply nonlinear $\mathcal{B}_{\kappa}$-embeddings to any situation involving connections among mathematical objects with different dimensions and degrees of freedom. Examples of these situations are provided by conformational changes and phase transitions in Statistical Mechanics, irreversible processes involving information loss (e.g. coarse-graining of the microscopic dynamics) and the holographic principle (which relates degrees of freedom of quantum field theories in different dimensions).  

Another specific example of a situation involving objects with different dimensions is provided by the quest for a unified theory of general relativity and quantum mechanics in the framework of, e.g. supergravity. These proceed by extending the 4-dimensional metric of pure gravity to higher dimensions \cite{Krasnov, overduin, bailin, Witten81,duff86}. Although the observable universe is described by a 4-dimensional metric tensor, this latter object needs to have 10 or 11 dimensions (depending on whether supergravity arises as the low energy limit of string theory or M-theory, respectively \cite{Krasnov}) in order to consistently accomodate the gauge groups describing the standard model of particle physics \cite{Witten81, Beck}. Since there are only 4 observable dimensions of spacetime, warped models for the metric are considered to be able to change the dimensionality of the metric \cite{RS1, Becker}. With help of appropriate nonlinear  $\mathcal{B}_{\kappa}$-embeddings we point out how new warped models and a $\kappa$-deformed formalism of gravity can be constructed

Nonlinear $\mathcal{B}_{\kappa}$-embeddings can also be applied to dynamical systems, as cellular automata (CAs) \cite{Wolfram, DeutschBOOK, Tokihiro1, Ilachinski, Adamatzky, Wuensche, VGM1, semipredo,VGM2,VGM3,JPHYSA,diagrammar} and coupled map lattices (CMLs) \cite{Kaneko1, Kaneko2, Kapral, KanekoBOOK, Bunimovich2, Bunimovich3,JPHYSA}.  These models of complex physical systems are popular in biophysics \cite{DeutschBOOK}, have given rise to novel approaches to quantum mechanics \cite{semipredo,diagrammar,thooft,VGMquantum} and have been conjectured to play a crucial role in unified field theories, giving rise to the concept of chaotic strings in the framework of stochastic quantization \cite{Beck}. However, natural physical systems display a great deal of variability and their evolution departs from the specification of a few rigid, deterministic rules perfectly operating on finite amounts of information. It is, therefore, interesting to study how these models can be embedded in more sophisticated ones \cite{Chate1, Chate2, Tokihiro1,Bagnoli,Just,JPHYSA}. In a previous recent work \cite{JPHYSA}, we have presented a general mechanism that allows any CA to be embedded in a CML in terms of a control parameter $\kappa$ that governs the embedding. We display here new constructions and we show how these yield nonlinear $\mathcal{B}_{\kappa}$-embeddings that are able to connect CAs to nonlinear partial differential equations (PDEs) and derive from these connections certain nonlinear diffusion equations. Furthermore, we construct $\mathcal{B}_{\kappa}$-embeddings ($\kappa$-deformed structures) corresponding to CMLs that are able to glue together several different CAs. In similar ways, CMLs can be glued together to form more complicated structures. The mathematical methods presented here may be of interest in biophysics (dynamics of multicellular ensembles) and in fundamental physics (extended formalisms of gravity and the embedding of different unified theories of physics related by dualities). Quite interestingly, a certain class of CMLs have been used to simulate quantum field theories on an appropriate scaling limit \cite{Beck, Beck2} and it has been shown that there are 6 different such unified theories in terms of chaotic strings that are somehow analogous to the six different models of a string considered in string theory and which are embedded in M-theory \cite{Beck, Beck3}.

The outline of this article is as follows. First, in Section \ref{embeddings} we introduce the method to construct the nonlinear $\mathcal{B}_{\kappa}$-embeddings involved in connecting structures with different dimensions (thus generalizing the concept of vectors and matrices). In Section  \ref{examplesec} we give several specific examples of nonlinear embeddings. In Section \ref{compact} we apply nonlinear embeddings to the problem of connecting a 4-dimensional and a 11-dimensional metric tensors, as those found in theories of supergravity, at appropriate limits of the $\kappa$ parameter. In this application, $\kappa$ is related to the characteristic scale at which spacetime is probed when compared to the Planck length. In Section \ref{CA2PDE} we construct $\mathcal{B}_{\kappa}$-embeddings that have CAs in the $\kappa \to 0$ limit and we show how they are connected in the asymptotic limit $\kappa$ large to certain nonlinear partial differential equations. We derive through this method a class of nonlinear diffusion equations, and discuss the parameter values that lead to linear diffusion equations. Finally, in Section \ref{CMLcon} we show how two arbitrary CAs in rule space can be connected in the $\kappa \to 0$ and $\kappa \to \infty$ limits by means of appropriate nonlinear $\mathcal{B}_{\kappa}$-embeddings. We discuss some potential physical applications and present some conclusions.

\section{Nonlinear $\mathcal{B}_{\kappa}$-embeddings: connecting scalars, vectors and matrices} \label{embeddings}

The method to construct $\mathcal{B}_{\kappa}$-embeddings connecting objects with different dimensions begins by noting that the Kronecker delta $\delta_{nj}$ ($\delta_{nj}=1$ if $n=j$ and $\delta_{nj}=0$ otherwise) admits a simple representation in terms of the boxcar function \cite{VGM1, VGM2}
\begin{eqnarray}
\mathcal{B}(x,y)&\equiv&\frac{1}{2}\left(\frac{x+y}{|x+y|}-\frac{x-y}{|x-y|}\right)={\begin{cases} \text{sign}(y)&{\text{if }}|x| < |y|\\ \text{sign}(y)/2 &{\text{if }}|x|=|y|, y\ne 0 \\0& {\text{otherwise}} \end{cases}} \label{d1}
\end{eqnarray} 
where $x, y \in \mathbb{R}$. Indeed, we have
\begin{equation}
\delta_{nj}=\mathcal{B}\left(n-j,\frac{1}{2}\right) \label{d2}
\end{equation} 


We now note that, by means of convolution, any $N$-tuple $\mathbf{v}=(v_0, v_1,\ldots, v_{N-1})$ can be written in terms of its components $v_{n}\in \mathbb{C}$ as
\begin{equation}
(\mathbf{v})_n \equiv v_n= \sum_{j=0}^{N-1}v_j\delta_{nj}, ~~~~~~~~ (n=0,1,\ldots, N-1) \label{vector}
\end{equation}
where $n$ is a free index labelling the component. 

We note that Eq. (\ref{vector}) can equivalently be written by using Eq. (\ref{d2}) as
\begin{equation}
(\mathbf{v})_n \equiv v_n= \sum_{j=0}^{N-1}v_{j}\mathcal{B}\left(n-j,\frac{1}{2} \right), ~~~~~~~~ (n=0,1,\ldots, N-1) \label{vector2}
\end{equation}
It is straightforward to prove that the definition of a vector given by the r.h.s. of Eq. (\ref{vector2}) is indeed consistent with that of an element of a vector space $V$ of $N$ dimensions. All eight  following properties are satisfied:\\
\begin{quote}
1. Associative property, for any three vectors $\mathbf{u}$, $\mathbf{v}$ and $\mathbf{w}$ $\in V$
\begin{eqnarray}
\left[\mathbf{u}+(\mathbf{v}+\mathbf{w})\right]_n&=&\sum_{j=0}^{N-1}\left[u_j+(v_{j}+w_j)\right]\mathcal{B}\left(n-j,\frac{1}{2} \right) ~~~~~~~~~ (n=0,1,\ldots, N-1) \nonumber \\
&=&\sum_{j=0}^{N-1}\left[(u_j+v_{j})+w_j\right]\mathcal{B}\left(n-j,\frac{1}{2} \right)=\left[(\mathbf{u}+\mathbf{v})+\mathbf{w}\right]_n  \nonumber
\end{eqnarray}
2. Conmutativity of addition, for any two vectors $\mathbf{u}$ and $\mathbf{v}$ $\in V$
\begin{eqnarray}
(\mathbf{u}+\mathbf{v})_n&=&\sum_{j=0}^{N-1}\left(u_j+v_{j}\right)\mathcal{B}\left(n-j,\frac{1}{2} \right) \qquad \qquad \qquad \qquad  (n=0,1,\ldots, N-1) \nonumber \\
&=& \sum_{j=0}^{N-1}\left(v_j+u_{j}\right)\mathcal{B}\left(n-j,\frac{1}{2} \right)=(\mathbf{v}+\mathbf{u})_n  \nonumber 
\end{eqnarray}
3. Identity of addition: there exists $\mathbf{0}$, $\mathbf{0}_n \equiv \sum_{j=0}^{N-1}0\mathcal{B}\left(n-j,\frac{1}{2} \right)$ such that for any vector $\mathbf{u}$ $\in V$
\begin{eqnarray}
(\mathbf{u}+\mathbf{0})_n&=&\sum_{j=0}^{N-1}\left(u_j+0\right)\mathcal{B}\left(n-j,\frac{1}{2} \right)=\sum_{j=0}^{N-1}u_{j}\mathcal{B}\left(n-j,\frac{1}{2} \right)=(\mathbf{u})_n \nonumber
\end{eqnarray}
4. Inverse elements of addition: for any vector $\mathbf{u}$, $(\mathbf{u})_n\equiv \sum_{j=0}^{N-1}u_j\mathcal{B}\left(n-j,\frac{1}{2} \right)$ there exists a vector $-\mathbf{u}$ with components $(-\mathbf{u})_n \equiv \sum_{j=0}^{N-1}(-u_j)\mathcal{B}\left(n-j,\frac{1}{2} \right)$ $\in V$ such that  
\begin{eqnarray}
[\mathbf{u}+(-\mathbf{u})]_n&=&\sum_{j=0}^{N-1}\left(u_j+(-u_j)\right)\mathcal{B}\left(n-j,\frac{1}{2} \right)=\sum_{j=0}^{N-1}0\mathcal{B}\left(n-j,\frac{1}{2} \right)=(\mathbf{0})_n \nonumber
\end{eqnarray}
for $n=0,1,\ldots N-1$.\\
5. Compatibility of scalar multiplication with field multiplication. For any two $a$, $b$ scalars in a field $\mathbb{F}$, we have 
\begin{eqnarray}
a(b\mathbf{u})_n&=&a\sum_{j=0}^{N-1}(bu_j)\mathcal{B}\left(n-j,\frac{1}{2} \right)=\sum_{j=0}^{N-1}(ab)u_j\mathcal{B}\left(n-j,\frac{1}{2} \right)=(ab)(\mathbf{u})_n \nonumber
\end{eqnarray}
6. Identity element of scalar multiplication. Let 1 denote the multiplicative unit in the field $\mathbb{F}$. Then, we have 
\begin{eqnarray}
1(\mathbf{u})_n&=&\sum_{j=0}^{N-1}1u_j\mathcal{B}\left(n-j,\frac{1}{2} \right)=\sum_{j=0}^{N-1}u_j\mathcal{B}\left(n-j,\frac{1}{2} \right)=(\mathbf{u})_n \nonumber
\end{eqnarray}
for $n=0,1,\ldots N-1$.\\
7. Distributivity of scalar multiplication with respect to vector addition. For any two vectors $\mathbf{u}$, $\mathbf{v}$ and a scalar $a\in \mathbb{F}$ we have 
\begin{eqnarray}
a(\mathbf{u}+\mathbf{v})_n&=&\sum_{j=0}^{N-1}a(u_j+v_j)\mathcal{B}\left(n-j,\frac{1}{2} \right)=\sum_{j=0}^{N-1}(au_j+av_j)\mathcal{B}\left(n-j,\frac{1}{2} \right)=a(\mathbf{u})_n+a(\mathbf{v})_n \nonumber
\end{eqnarray}
for $n=0,1,\ldots N-1$.\\
8. Distributivity of scalar multiplication with respect to field addition. For any vector $\mathbf{u}$, $\mathbf{v}$ and any two scalars $a,b\in \mathbb{F}$ we have 
\begin{eqnarray}
(a+b)(\mathbf{u})_n&=&\sum_{j=0}^{N-1}(a+b)u_j\mathcal{B}\left(n-j,\frac{1}{2} \right)=\sum_{j=0}^{N-1}(au_j+bu_j)\mathcal{B}\left(n-j,\frac{1}{2} \right) \nonumber \\
&=&a(\mathbf{u})_n+b(\mathbf{u})_n ~~~~~~~~~ (n=0,1,\ldots, N-1) \nonumber
\end{eqnarray}
with $n=0,1,\ldots N-1$.\\
\end{quote}
The Kronecker product $\mathbf{v}\otimes \mathbf{w}$ of two vectors $\mathbf{v}$ and $\mathbf{w}$ with dimensions $N$ and $M$, respectively, leads to a matrix of size $N \times M$ with two free indices 
\begin{equation}
(\mathbf{v}\otimes \mathbf{w})_{nm} \equiv v_n w_m=\sum_{j=0}^{N-1}\sum_{k=0}^{M-1}v_{j}w_{k}\mathcal{B}\left(n-j, \frac{1}{2}\right)\mathcal{B}\left(m-k, \frac{1}{2}\right) \label{outpro}
\end{equation} 
with $n=0,1,\ldots N-1$ and $m=0,1,\ldots M-1$.
 
If $M=N$ the inner product $\left< \ldots \right>$ of two vectors $\mathbf{v}$ and $\mathbf{w}$ is given by
\begin{equation}
\left<\mathbf{v}, \mathbf{w}\right> \equiv \sum_{n=0}^{N-1}v_{n}w_{n}
\end{equation} 
and can be obtained from the outer product by contracting the free indices
\begin{eqnarray}
\left<\mathbf{v}, \mathbf{w}\right> &=& \sum_{n=0}^{N-1}\sum_{m=0}^{N-1}\mathcal{B}\left(n-m, \frac{1}{2}\right)   \sum_{j=0}^{N-1}\sum_{k=0}^{N-1}v_{j}w_{k}\mathcal{B}\left(n-j, \frac{1}{2}\right)\mathcal{B}\left(m-k, \frac{1}{2}\right) \label{outin}
\end{eqnarray} 

We define a ket vector $\ket{\psi}$ over an $N$-dimensional vector space in terms of $\mathcal{B}$-functions as
\begin{equation}
\ket{\psi}\equiv \sum_{j=0}^{N-1}\psi_{j}\mathcal{B}\left(n-j, \frac{1}{2}\right)
\end{equation}
where the $\psi_{j}$ are complex numbers. A bra vector is defined as
\begin{equation}
\bra{\psi}\equiv \sum_{j=0}^{N-1}\overline{\psi}_{j}\mathcal{B}\left(n-j, \frac{1}{2}\right)
\end{equation}
where the overline denotes complex conjugation. The inner product of a bra and a ket is then obtained as
\begin{equation}
\braket{\phi|\psi}\equiv  \sum_{n=0}^{N-1}\sum_{m=0}^{N-1}\mathcal{B}\left(n-m, \frac{1}{2}\right)\sum_{j=0}^{N-1}\sum_{k=0}^{N-1}\overline{\phi}_{j}\psi_{k} \mathcal{B}\left(n-j, \frac{1}{2}\right) \mathcal{B}\left(m-k, \frac{1}{2}\right)=\sum_{n=0}^{N-1}\overline{\phi}_{n}\psi_{n}
\end{equation}
Let $a$ and $b$ be complex numbers. We clearly have the following properties
\begin{quote}
1. $\ket{a\psi}=a\ket{\psi}$ \\
2. $\bra{a\psi}=\overline{a}\bra{\psi}$ \\
3. $\braket{a\phi_{1}+b\phi_{2}|\psi}=\overline{a}\braket{\phi_{1}|\psi}+\overline{b}\braket{\phi_{2}|\psi}$ \\
4. $\braket{\phi| a\psi_{1}+b\psi_{2}}=a\braket{\phi|\psi_1}+b\braket{\phi|\psi_{2}}$ \\
5. $\braket{\psi| \psi} \ge 0$ \\
6. Let $d(\phi,\psi) \equiv \sqrt{\braket{\psi-\phi |\psi-\phi } }$. Then:
\begin{quote}
6.1. $d(\psi,\psi)=0$;\\
6.2. $d(\phi,\psi)=d(\psi,\phi)$;\\
6.3. $d(\phi,\psi)\leq d(\phi,\eta)+d(\eta,\psi)$
\end{quote} 
\end{quote}
Because of these properties, the vector space with inner product defined above is a Hilbert space.

An $N\times N$ matrix $\mathbf{A}$ can be written in terms of its elements $A_{nm}$ as
\begin{equation}
(\mathbf{A})_{nm}\equiv A_{nm} = \sum_{j=0}^{N-1}\sum_{k=0}^{N-1}A_{jk}\mathcal{B}\left(n-j,\frac{1}{2} \right)\mathcal{B}\left(m-k,\frac{1}{2} \right) ~~~~ (n, m=0,1,\ldots N-1)  \label{matrix}
\end{equation}

We note that an $N$-dimensional matrix can equivalently be written in a more compact form as a vector of indexed $N$-tuples. If we define
\begin{equation}
A_{h}=A_{k+Nj}\equiv A_{jk} \qquad ~~~~ (j, k=0,1,\ldots N-1)
\end{equation}
where $h\equiv k+Nj$ then Eq. (\ref{matrix}) becomes
\begin{equation}
(\mathbf{A})_{nm}=\sum_{h=0}^{N^2-1}A_{h}\mathcal{B}\left(h-m-Nn,\frac{1}{2} \right) \qquad ~~~~ (n, m=0,1,\ldots N-1) \label{matrix2}
\end{equation}
Note that there are two free indices $m$ and $n$ in this expression and that there are $N^2$ coefficients $A_{h}$.

The $\mathcal{B}$-function (boxcar function) is a suitable representation of the Kronecker delta when its first argument $x$ is an integer and its second argument $y$ has value $|y| \le \frac{1}{2}$. The crucial interest of this representation of the Kronecker delta is that it can be easily embedded in the real numbers by means of a one-parameter deformation function $\mathcal{B}_{\kappa}(x,y)$ which (pointwise) converges to $\mathcal{B}(x,y)$ as $\kappa \to 0$ and (uniformly) converges to 0 as $\kappa \to \infty$ \cite{JPHYSA,homotopon}
\begin{equation}
\mathcal{B}_{\kappa}(x,y)\equiv \frac{1}{2}\left[ 
\tanh\left(\frac{x+y}{\kappa} \right)-\tanh\left(\frac{x-y}{\kappa} \right)
\right] \label{bkappa}
\end{equation}
with $x,y\in \mathbb{R}$. It is straightforward to observe that
\begin{eqnarray}
\mathcal{B}_{\kappa}\left(-x, y \right)&=&\mathcal{B}_{\kappa}\left(x, y \right) \label{pari1}\\
\mathcal{B}_{\kappa}\left(x, -y \right)&=&-\mathcal{B}_{\kappa}\left(x, y \right) \label{pari2}\\
\mathcal{B}_{-\kappa}\left(x, y \right)&=&-\mathcal{B}_{\kappa}\left(x, y \right) \label{pari3}
\end{eqnarray}
and to check all following identities
\begin{eqnarray}
0< \mathcal{B}_{\kappa}\left(x, y \right) &< & \text{sign}\ y \qquad \forall{\kappa} \qquad \label{lim00} \\
0< \frac{\mathcal{B}_{\kappa}\left(x, y\right)}{\mathcal{B}_{\kappa}\left(0, y \right)} &< & 1 \qquad \forall{\kappa} \qquad \label{lim11} \\
\lim_{\kappa \to 0}\mathcal{B}_{\kappa}\left(x,y\right)&=&  \mathcal{B}\left(x,y\right) \label{lim1}  \\
\lim_{\kappa \to 0}\frac{\mathcal{B}_{\kappa}\left(x, y\right)}{\mathcal{B}_{\kappa}\left(0, y\right)}&=&\frac{\mathcal{B}\left(x, y\right)}{\mathcal{B}\left(0, y\right)}=\mathcal{B}\left(x,|y|\right) \label{lim1b} \\
\lim_{\kappa \to \infty}\mathcal{B}_{\kappa}\left(x, y\right)&=& 0  \label{lim2} \\
\lim_{\kappa \to \infty}\frac{\mathcal{B}_{\kappa}\left(x,y\right)}{\mathcal{B}_{\kappa}\left(0, y \right)}&=&1 \label{lim3}
\end{eqnarray}
by noting that, after some manipulations, we have
\begin{eqnarray}
\mathcal{B}_{\kappa}\left(x, y\right)&=&\frac{e^{2y/\kappa}-e^{-2y/\kappa}}{e^{2y/\kappa}+e^{2x/\kappa}+e^{-2x/\kappa}+e^{-2y/\kappa}} \\
\frac{\mathcal{B}_{\kappa}\left(x, y\right)}{\mathcal{B}_{\kappa}\left(0, y\right)}&=&\frac{e^{2y/\kappa}+2+e^{-2y/\kappa}}{e^{2y/\kappa}+e^{2x/\kappa}+e^{-2x/\kappa}+e^{-2y/\kappa}}
\end{eqnarray} 
We also observe the fact that, while $\mathcal{B}_{\kappa}\left(x, y \right)$ is an even function of $x$ and an odd function of $\kappa$ and $y$, the ratio $\mathcal{B}_{\kappa}\left(x, y \right)/\mathcal{B}_{\kappa}\left(0, y \right)$ is an even function of all $x$, $y$ and $\kappa$. We note, furthermore, that \cite{JPHYSA}
\begin{eqnarray}
\mathcal{B}_{\kappa}(x,y+z)&=&\mathcal{B}_{\kappa}(x+y,z)+\mathcal{B}_{\kappa}(x-z,y)  \label{iden1} \\
\sum_{k=0}^{N-1}\mathcal{B}_{\kappa}(x-2ky,y)&=&\mathcal{B}_{\kappa}(x-(N-1)y,Ny) \label{iden2}
\end{eqnarray}
Interestingly, we have the following expansions
\begin{eqnarray}
e^{x/\kappa}&=&\frac{1+\mathcal{B}_{\kappa}\left(0,\frac{x}{2}\right)}{1-\mathcal{B}_{\kappa}\left(0,\frac{x}{2}\right)}=\left[1+\mathcal{B}_{\kappa}\left(0,\frac{x}{2}\right)\right]\sum_{j=0}^{\infty}\left[\mathcal{B}_{\kappa}\left(0,\frac{x}{2}\right)\right]^j \nonumber \\
&=&1+2\sum_{j=1}^{\infty}\left[\mathcal{B}_{\kappa}\left(0,\frac{x}{2}\right)\right]^j \label{eexp} \\
\cosh \frac{x}{\kappa}&=&1+2\sum_{j=1}^{\infty}\left[\mathcal{B}_{\kappa}\left(0,\frac{x}{2}\right)\right]^{2j} \label{coshexp} \\
\sinh \frac{x}{\kappa}&=&2\sum_{j=1}^{\infty}\left[\mathcal{B}_{\kappa}\left(0,\frac{x}{2}\right)\right]^{2j-1} \label{coshexp}
\end{eqnarray} 

The $\mathcal{B}_{\kappa}$-function is an infinitely differentiable function of $x, y$ and $\kappa$. The $n$-th derivatives are given by 
\begin{eqnarray}
\frac{\partial^n}{\partial x^{n}}\mathcal{B}_{\kappa}(x,y)&=&(2\kappa)^{n}\sum_{j=0}^{n-1}\left\langle {n \atop j}\right\rangle \left[\frac{e^{2(x+y)(1+j)/\kappa}}{\left(1+e^{2(x+y)/\kappa}\right)^{n+1}}-\frac{e^{2(x-y)(1+j)/\kappa}}{\left(1+e^{2(x-y)/\kappa}\right)^{n+1}} \right] \label{dnx} \\
\frac{\partial^n}{\partial y^{n}}\mathcal{B}_{\kappa}(x,y)&=&(2\kappa)^{n}\sum_{j=0}^{n-1}\left\langle {n \atop j}\right\rangle \left[\frac{e^{2(x+y)(1+j)/\kappa}}{\left(1+e^{2(x+y)/\kappa}\right)^{n+1}}+\frac{e^{-2(x-y)(1+j)/\kappa}}{\left(1+e^{-2(x-y)/\kappa}\right)^{n+1}} \right] \label{dny} \\
\frac{\partial^n}{\partial (1/\kappa)^{n}}\mathcal{B}_{\kappa}(x,y)&=&2^{n}\sum_{j=0}^{n-1}\left\langle {n \atop j}\right\rangle \left[\frac{(x+y)^{n}e^{2(x+y)(1+j)/\kappa}}{\left(1+e^{2(x+y)/\kappa}\right)^{n+1}}-\frac{(x-y)^{n}e^{2(x-y)(1+j)/\kappa}}{\left(1+e^{2(x-y)/\kappa}\right)^{n+1}} \right] \label{dnkappa}
\end{eqnarray}
where we have introduced the Eulerian number
\begin{equation}
\left\langle {n \atop j}\right\rangle \equiv \sum_{m=0}^{j+1}(-1)^{m}{n+1 \choose m}(j+1-m)^{n}
\end{equation}

We also have 
\begin{equation}
\int_{a}^{b} dx \mathcal{B}_{\kappa}\left(x, y\right)=\frac{\kappa}{2}\ln \left[\frac{\cosh \frac{b+y}{\kappa}}{\cosh \frac{b-y}{\kappa}} \frac{\cosh \frac{a-y}{\kappa}}{\cosh \frac{a+y}{\kappa}} \right]
\end{equation}
Therefore, independently of $\kappa$,
\begin{equation}
\frac{1}{2y}\int_{-\infty}^{\infty} dx \mathcal{B}_{\kappa}\left(x, y\right)=1
\end{equation}

For $\kappa$ sufficiently large ($\kappa > \frac{2(|x|+|y|)}{\pi}$) the hyperbolic tangents in the definition of the $\mathcal{B}_{\kappa}$-function can be expanded in their convergent Maclaurin series, and we have 
\begin{eqnarray}
\mathcal{B}_{\kappa}\left(x, y\right)&=&\frac{1}{2}\sum_{j=1}^{\infty}\frac{2^{2j}(2^{2j}-1)B_{2j}}{(2j)!\kappa^{2j-1}}\left[\left(x+y \right)^{2j-1}-\left(x-y\right)^{2j-1}\right] \nonumber \\
&=&  \frac{1}{2}\sum_{j=1}^{\infty}\frac{2^{2j}(2^{2j}-1)B_{2j}}{(2j)!\kappa^{2j-1}}
\sum_{h=0}^{2j-1}{2j-1 \choose h}x^{2j-1-h}(1-(-1)^{h})y^{h} \nonumber \\
&=&  \sum_{j=1}^{\infty}\frac{2^{2j}(2^{2j}-1)B_{2j}}{(2j)! \kappa^{2j-1}}
\sum_{h=1}^{j}{2j-1 \choose 2h-1}x^{2(j-h)}y^{2h-1}   \nonumber \\
&=& \mathcal{B}_{\kappa}\left(0, y\right)+\sum_{j=2}^{\infty}\frac{2^{2j}(2^{2j}-1)B_{2j}}{(2j)! \kappa^{2j-1}}
\sum_{h=1}^{j-1}{2j-1 \choose 2h-1}y^{2h-1} x^{2(j-h)} \nonumber \\ 
&=&\frac{y}{\kappa}-\frac{y^3+3x^2y}{3\kappa^3}+O\left(\kappa^{-5}\right) \label{Bernoultheo1} \\
\frac{\mathcal{B}_{\kappa}\left(x, y\right)}{\mathcal{B}_{\kappa}\left(0, y\right)}&=&1+\frac{1}{\mathcal{B}_{\kappa}\left(0, y\right)}\sum_{j=2}^{\infty}\frac{2^{2j}(2^{2j}-1)B_{2j}}{(2j)! \kappa^{2j-1}} 
\sum_{h=1}^{j-1}{2j-1 \choose 2h-1}y^{2h-1} x^{2(j-h)} \nonumber \\ 
&=&1-\frac{x^2}{\kappa^2}+O\left(\kappa^{-4}\right) \label{Bernoultheo2} 
\end{eqnarray}
where the $B_{2m}$ denote the even Bernoulli numbers: $B_{0}=1$, $B_{2}=\frac{1}{6}$, $B_{4}=-\frac{1}{30}$ $B_{6}=\frac{1}{42}$, etc. In the above expression we have also used that
\begin{eqnarray}
\mathcal{B}_{\kappa}\left(0, y\right)&=&  \sum_{j=1}^{\infty}\frac{2^{2j}(2^{2j}-1)B_{2j}y^{2j-1}}{(2j)! \kappa^{2j-1}}=\tanh \frac{y}{\kappa} \label{Bernoultheo2} 
\end{eqnarray}

The main trick introduced in this manuscript is to replace any $\mathcal{B}$-function in Eqs. (\ref{vector2}) or (\ref{matrix}), e.g. $\mathcal{B}\left(n-j,\frac{1}{2} \right)$,
by a $\kappa$-deformed counterpart with the form either $\mathcal{B}_{\kappa}\left(n-j,\frac{1}{2} \right)$ or $\mathcal{B}_{\kappa}\left(n-j,\frac{1}{2} \right)/\mathcal{B}_{\kappa}\left(0,\frac{1}{2} \right)$ depending on the application under consideration. 
We shall call the first kind of replacement Mode I and the latter Mode II. If we simply replace all $\mathcal{B}$-functions following Mode I we obtain from Eqs. (\ref{vector2}) or (\ref{matrix}), respectively,
\begin{eqnarray}
v_{\kappa}^{(I)}&\equiv& \sum_{j=0}^{N-1}v_{j}\mathcal{B}_{\kappa}\left(n-j,\frac{1}{2} \right) \label{vector3} \\
A_{\kappa}^{(I)}&\equiv&\sum_{j=0}^{N-1}\sum_{k=0}^{N-1}A_{jk}\mathcal{B}_{\kappa}\left(n-j,\frac{1}{2} \right)\mathcal{B}_{\kappa}\left(m-k,\frac{1}{2} \right) \label{matrix3}
\end{eqnarray}
These nonlinear $\mathcal{B}_{\kappa}$-embeddings constitute $\kappa$-deformed structures that generalize those of a vector and of a matrix respectively as follows: In the limit $\kappa \to 0$, by using Eq. (\ref{lim1}) in Eqs. (\ref{vector3}) and (\ref{matrix3}), we regain Eqs. (\ref{vector2}) and (\ref{matrix}), respectively. However, in the limit $\kappa \to \infty$, from Eq. (\ref{lim2}), we obtain $\lim_{\kappa\to \infty}v_{\kappa}^{(I)}=0$ and $\lim_{\kappa\to \infty}A_{\kappa}^{(I)}=0$. 

If, instead, we replace all $\mathcal{B}$-functions following Mode II we obtain from Eqs. (\ref{vector2}) or (\ref{matrix}), respectively
\begin{eqnarray}
v_{\kappa}^{(II)}&\equiv&\sum_{j=0}^{N-1}v_{j}\frac{\mathcal{B}_{\kappa}\left(n-j,\frac{1}{2} \right)}{\mathcal{B}_{\kappa}\left(0,\frac{1}{2}\right)} \label{vector4} \\
A_{\kappa}^{(II)}&\equiv&\sum_{j=0}^{N-1}\sum_{k=0}^{N-1}A_{jk}\frac{\mathcal{B}_{\kappa}\left(n-j,\frac{1}{2} \right)}{\mathcal{B}_{\kappa}\left(0,\frac{1}{2}\right)}\frac{\mathcal{B}_{\kappa}\left(m-k,\frac{1}{2} \right)}{\mathcal{B}_{\kappa}\left(0,\frac{1}{2}\right)} \label{matrix4}
\end{eqnarray}
In the limit $\kappa \to 0$, by using Eq. (\ref{lim1}) in Eqs. (\ref{vector3}) and (\ref{matrix3}), we regain Eqs. (\ref{vector2}) and (\ref{matrix}), respectively. However, in the limit $\kappa \to \infty$, from Eq. (\ref{lim3}), we now obtain 
\begin{eqnarray}
\lim_{\kappa\to \infty}v_{\kappa}^{(II)}&=&\sum_{j=0}^{N-1}v_{j}\equiv v \label{scalarsum} \\
\lim_{\kappa\to \infty}A_{\kappa}^{(II)}&=&\sum_{j=0}^{N-1}\sum_{k=0}^{N-1}A_{jk}\equiv A \label{matrixsum}
\end{eqnarray}
i.e., the vector and the matrix collapse, respectively, to the scalars $v$ and $A$ formed by summing over all their entries.

 Note that, in both modes, the dimensionality of the mathematical object is reduced from $N$ to $0$ as $\kappa$ is varied from $0$ to $\infty$ when all $\mathcal{B}$-functions are being replaced. If one chooses to replace only a finite subset of the $\mathcal{B}$-functions and combines both modes of replacement, the dimensionality of the object can be tuned to any integer value between 0 and $N$.
Which replacement mode to choose and which entries of the matrix are to be deformed by the $\kappa$ parameter depend on the application at hand. For example, let us imagine that we want to construct a nonlinear $\mathcal{B}_{\kappa}$-embedding that connects any arbitrary operator $\mathbf{A}$ with its trace $\text{Tr} \mathbf{A}$. Then we begin by noting that we can equivalently write $\mathbf{A}$ in terms of its elements as
\begin{eqnarray}
(\mathbf{A})_{nm}&=&\sum_{j=0}^{N-1}\sum_{k=0}^{N-1}A_{jk}\mathcal{B}\left(n-j,\frac{1}{2} \right)\mathcal{B}\left(m-k,\frac{1}{2} \right) \label{matrixTr} \\
&=&\sum_{j=0}^{N-1}A_{jj}\mathcal{B}\left(n-j,\frac{1}{2} \right)\mathcal{B}\left(m-j,\frac{1}{2} \right)+\sum_{j=0}^{N-1}\sum_{k\ne j}A_{jk}\mathcal{B}\left(n-j,\frac{1}{2} \right)\mathcal{B}\left(m-k,\frac{1}{2} \right) \nonumber
\end{eqnarray}
From this, we can combine replacement Modes I and II to construct a nonlinear $\mathcal{B}_{\kappa}$-embedding
\begin{eqnarray}
A_{\kappa}&=&\sum_{j=0}^{N-1}A_{jj}\frac{\mathcal{B}_{\kappa}\left(n-j,\frac{1}{2} \right)}{\mathcal{B}_{\kappa}\left(0,\frac{1}{2}\right)}\frac{\mathcal{B}_{\kappa}\left(m-j,\frac{1}{2} \right)}{\mathcal{B}_{\kappa}\left(0,\frac{1}{2}\right)}+\sum_{j=0}^{N-1}\sum_{k\ne j}A_{jk}\mathcal{B}_{\kappa}\left(n-j,\frac{1}{2} \right)\mathcal{B}_{\kappa}\left(m-k,\frac{1}{2} \right) \label{matrixTrK}
\end{eqnarray}
In the limit $\kappa \to 0$, by using Eq. (\ref{lim1}) in Eq. (\ref{matrixTrK}), we regain Eq. (\ref{matrixTr}). However, in the limit $\kappa \to \infty$, from Eq. (\ref{lim3}), we now obtain 
\begin{eqnarray}
\lim_{\kappa\to \infty}A_{\kappa}&=&\sum_{j=0}^{N-1}A_{jj}=\text{Tr} \textbf{A} \label{matrixsumTr}
\end{eqnarray}
In this way, the nonlinear $\mathcal{B}_{\kappa}$-embedding given by Eq. (\ref{matrixTr}) smoothly connects a matrix operator with its (scalar) trace.

We note that in any nonlinear $\mathcal{B}_{\kappa}$-embedding, the limits $\kappa \to 0$ and $\kappa \to \infty$ can be exchanged by making the transformation $\kappa \to 1/\kappa$. Thus, for any nonlinear $\mathcal{B}_{\kappa}$-embedding, there exists a `dual' structure obtained by making the latter transformation.

\section{Examples of nonlinear $\mathcal{B}_{\kappa}$-embeddings} \label{examplesec}

 We now give more specific examples of nonlinear $\mathcal{B}_{\kappa}$-embeddings. The following one is inspired in the process of cell division found in biological systems \cite{Maton}. We can model the space occupied by a cell by means of the function
\begin{equation}
\mathcal{C}_{0}(x,y)=\mathcal{B}_{0.1}\left(\sqrt{x^2+y^2},1 \right) \label{C0}
\end{equation}
which is approximately equal to one inside a circle of unit radius and zero outside. Note that $\kappa=0.1$ in the $\mathcal{B}_{\kappa}$-function indicates that the border that separates the interior of the circle from its outside is smooth \cite{fuzzypap}. We can take $\mathcal{C}_{0}(x,y)$ as a part at $\kappa \approx 0$ of an embedding so that at the opposite limit, $\kappa \to \infty$, we have rather two cells with the centers displaced a certain distance along the $x$ axis
\begin{equation}
\mathcal{C}_{\infty}(x,y)=\mathcal{B}_{0.1}\left(\sqrt{\left(x-\frac{6}{5}\right)^2+y^2},1 \right)+\mathcal{B}_{0.1}\left(\sqrt{\left(x+\frac{6}{5}\right)^2+y^2},1 \right) \label{Cinf}
\end{equation}
We can now construct a $\mathcal{B}_{\kappa}$-embedding interpolating between these two limiting cases, introducing replacement modes I and II as
\begin{eqnarray}
\mathcal{C}_{\kappa}(x,y)&=&\mathcal{B}_{0.1}\left(\sqrt{x^2+y^2},1 \right)\mathcal{B}_{\kappa^3}\left(0,\frac{1}{2} \right) \label{citokin} \\
&&+\left[\mathcal{B}_{0.1}\left(\sqrt{\left(x-\frac{6\tanh \kappa}{5}\right)^2+y^2},1 \right)+\mathcal{B}_{0.1}\left(\sqrt{\left(x+\frac{6 \tanh \kappa}{5}\right)^2+y^2},1 \right)\right]\times \nonumber \\
&&\times \frac{\mathcal{B}_{\kappa}\left(1,\frac{1}{2} \right)}{\mathcal{B}_{\kappa}\left(0,\frac{1}{2} \right)} \nonumber
\end{eqnarray}
From this expression we clearly obtain
\begin{eqnarray}
\lim_{\kappa\to 0} \mathcal{C}_{\kappa}(x,y)=\mathcal{C}_{0}(x,y)\label{Ckappa0} \\
\lim_{\kappa\to \infty} \mathcal{C}_{\kappa}(x,y)=\mathcal{C}_{\infty}(x,y)\label{Ckappainf}
\end{eqnarray}

\begin{figure}
\includegraphics[width=0.6 \textwidth]{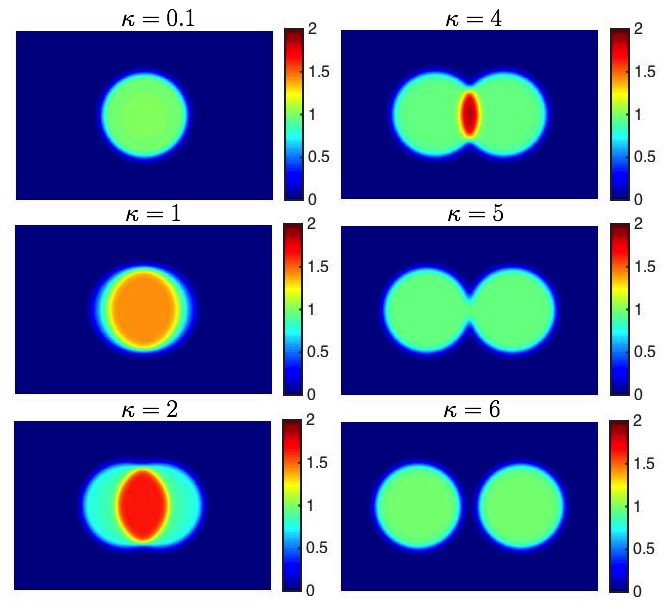}
\caption{\scriptsize{ Plot of the $\mathcal{C}_{\kappa}(x,y)$, given by Eq. (\ref{citokin}) in the plane $x\in [-3,3]$, $y\in [-2,2]$ for the values of $\kappa$ indicated over the panels.}} \label{mitos}
\end{figure}

In Fig. \ref{mitos}, the function $\mathcal{C}_{\kappa}(x,y)$, given by Eq. (\ref{citokin}), is represented in the plane for six different values of the parameter $\kappa$ indicated over the panels. We observe that for $\kappa=0.1$ the limit $\kappa \approx 0$ of the embedding, Eq. (\ref{C0}), is approached, while for $\kappa =6$ we obtain the limit $\kappa \to \infty$ of the embedding, Eq. (\ref{Cinf}). For intermediate $\kappa\in \mathbb{R}$ values we obtain a continuous, smooth transition that mimics the process of cell division (cytokinesis) found in many biological cells.

\begin{figure}
\includegraphics[width=0.6 \textwidth]{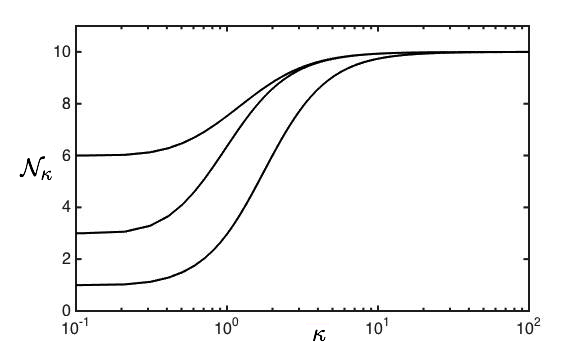}
\caption{\scriptsize{ The embedding $\mathcal{N}_{\kappa}$ given by Eq. (\ref{nkappa}). At large $\kappa$, the embedding collapses in a scalar value and we have $\mathcal{N}_{\kappa}=10$. At low $\kappa$ different branches arise and the embedding contains more information, behaving as a vector as $\kappa \to 0$ where the branches are  independent.}} \label{embed}
\end{figure}

Let us now consider another example. The following embedding connects a $N$-tuple of natural numbers to their sum
\begin{equation}
\mathcal{N}_{\kappa}(n)\equiv \frac{\mathcal{B}_{\kappa}\left(n, \frac{1}{2} \right)}{\mathcal{B}_{\kappa}\left(0, \frac{1}{2} \right)}+3\frac{\mathcal{B}_{\kappa}\left(n-1, \frac{1}{2} \right)}{\mathcal{B}_{\kappa}\left(0, \frac{1}{2} \right)}+6\frac{\mathcal{B}_{\kappa}\left(n-2, \frac{1}{2} \right)}{\mathcal{B}_{\kappa}\left(0, \frac{1}{2} \right)} \label{nkappa}
\end{equation}
This structure is typical of the replacement Mode II described above, which interpolates between the limits
\begin{eqnarray}
\mathcal{N}_{0}(n)&=&\lim_{\kappa \to 0} \mathcal{N}_{\kappa}=\mathcal{B}\left(n, \frac{1}{2} \right)+3\mathcal{B}\left(n-1, \frac{1}{2} \right)+6\mathcal{B}\left(n-2, \frac{1}{2} \right) \nonumber \\
\mathcal{N}_{\infty}&=&\lim_{\kappa \to \infty} \mathcal{N}_{\kappa}=1+3+6=10 \nonumber
\end{eqnarray}
The embedding $\mathcal{N}_{\kappa}$ given by Eq. (\ref{nkappa}) is plotted in Fig. \ref{embed} as a function of $\kappa$. We note the difference in the limiting behaviors $\kappa \to 0$ and $\kappa \to \infty$. At the former limit, the embedding has different branches, each indexed by different values of the variable $n$. At the latter limit the embedding collapses to an scalar value insensitive to $n$.

Following any of the branches of the embedding, from $\kappa \approx 0$ to $\kappa \to \infty$, information is lost. If we think in $\kappa$ as equivalent to time, the embedding collapses to a fixed point at $\kappa \to \infty$ and information on the starting branch at $\kappa \approx 0$ cannot be recovered. Nonlinear $\mathcal{B}_{\kappa}$- embeddings can thus be used to model \emph{irreversible processes} found in natural physical systems. 

We note that, along the branches $n=0,1,2$ we obtain, from Eq. (\ref{nkappa}) 
\begin{eqnarray}
\mathcal{N}_{\kappa}(0)&=& 1+3\frac{\mathcal{B}_{\kappa}\left(1, \frac{1}{2} \right)}{\mathcal{B}_{\kappa}\left(0, \frac{1}{2} \right)}+6\frac{\mathcal{B}_{\kappa}\left(2, \frac{1}{2} \right)}{\mathcal{B}_{\kappa}\left(0, \frac{1}{2} \right)} \label{nkappa0} \\
\mathcal{N}_{\kappa}(1)&=& 3+7\frac{\mathcal{B}_{\kappa}\left(1, \frac{1}{2} \right)}{\mathcal{B}_{\kappa}\left(0, \frac{1}{2} \right)} \label{nkappa1}\\
\mathcal{N}_{\kappa}(2)&=& 6+3\frac{\mathcal{B}_{\kappa}\left(1, \frac{1}{2} \right)}{\mathcal{B}_{\kappa}\left(0, \frac{1}{2} \right)}+\frac{\mathcal{B}_{\kappa}\left(2, \frac{1}{2} \right)}{\mathcal{B}_{\kappa}\left(0, \frac{1}{2} \right)} \label{nkappa2}
\end{eqnarray}
where we have used Eq. (\ref{pari1}).

Let $p(x)$ denote the total number of unrestricted partitions of the natural number $x$. It is clear that $x=10$ is obtained from Eq. (\ref{nkappa}) at $\kappa \to \infty$ starting from parts $1,3$ and $6$ at $\kappa\to 0$, but we could have considered other parts as well (e.g. $1$, $2$, $2$ and $5$) that would yield a different embedding with the same limiting behavior at $\kappa \to \infty$. Indeed, there are $p(10)=42$ possible embeddings whose parts at $\kappa \to 0$ are consistent with the sum of the parts being equal to 10 at $\kappa \to \infty$. Let $x_0 \in \mathbb{N}$ be the part of the embedding at $\kappa \approx 0$ and $x_\infty \in \mathbb{N}$ the part at $\kappa \to \infty$. We can quantify the irreversibility in going from $\kappa \approx 0$ to $\kappa \to \infty$ by means of the entropy change
\begin{equation}
\Delta S_{x_0 \to x_{\infty}}=\ln \frac{p(x_{\infty})}{p(x_{0})} \ge 0 \label{entropy}
\end{equation}
For the embedding in Eq. (\ref{nkappa}) and Fig.\ref{embed}, we have, along each branch of the embedding
\begin{eqnarray}
\Delta S_{1 \to 10}&=&\ln \frac{p(10)}{p(1)}=\ln \frac{42}{1} =3.74 \nonumber \\
\Delta S_{3 \to 10}&=&\ln \frac{p(10)}{p(3)}=\ln \frac{42}{3} =2.64  \nonumber \\
\Delta S_{6 \to 10}&=&\ln \frac{p(10)}{p(6)}=\ln \frac{42}{3} =1.34  \nonumber
\end{eqnarray}
The above entropy of the embedding between natural-valued parts exactly matches the entropy of a quantum mechanical nonrelativistic string with fixed endpoints (see, eg. \cite{Zwiebach}, pp. 498-500). This shows that nonlinear $\mathcal{B}_{\kappa}$-embeddings between natural-valued parts may find application in string theory. 


We note that, 
\begin{equation}
\frac{\partial }{\partial \kappa} \left[\frac{\mathcal{B}_{\kappa}(x,y)}{\mathcal{B}_{\kappa}(0,y)}\right]=\frac{2 \sinh ^2\left(\frac{x}{\kappa }\right) \cosh ^2\left(\frac{y}{\kappa }\right) \text{sech}^2\left(\frac{x-y}{\kappa }\right) \text{sech}^2\left(\frac{x+y}{\kappa }\right) \left(x \coth \left(\frac{x}{\kappa }\right)-y \tanh \left(\frac{y}{\kappa }\right)\right)}{\kappa ^2}
\end{equation}
and, for any $n\in \mathbb{Z}$ and $y=1/2$,
\begin{eqnarray}
\lim_{\kappa \to 0}\frac{\partial }{\partial \kappa} \left[\frac{\mathcal{B}_{\kappa}\left(n,\frac{1}{2}\right)}{\mathcal{B}_{\kappa}\left(0,\frac{1}{2}\right)}\right]=\lim_{\kappa \to \infty}\frac{\partial }{\partial \kappa} \left[\frac{\mathcal{B}_{\kappa}\left(n,\frac{1}{2}\right)}{\mathcal{B}_{\kappa}\left(0,\frac{1}{2}\right)}\right]=0 \label{thefundamental}
\end{eqnarray}
This is a crucial property of all nonlinear $\mathcal{B}_{\kappa}$-embeddings obtained by means of replacement mode II. To understand it let us consider two functions $f(x)$ and $g(x)$ of a real variable $x$. We can homotopically connect both by the following embedding
\begin{equation}
h_{\kappa}(x)=f(x)+\frac{\mathcal{B}_{\kappa}\left(1,\frac{1}{2}\right)}{\mathcal{B}_{\kappa}\left(0,\frac{1}{2}\right)}\left(g(x)-f(x)\right)
\end{equation}
for which we have the limits
\begin{eqnarray}
\lim_{\kappa \to 0}h_{\kappa}(x)&=&f(x) \\
\lim_{\kappa \to \infty}h_{\kappa}(x)&=&g(x)
\end{eqnarray}
Because of Eq. (\ref{thefundamental}), we have
\begin{eqnarray}
\lim_{\kappa \to 0}\frac{\partial h_{\kappa}(x)}{\partial \kappa}=\lim_{\kappa \to \infty}\frac{\partial h_{\kappa}(x)}{\partial \kappa}  =0 \label{thefundamental2}
\end{eqnarray}
The functions $f(x)$ and $g(x)$  are homotopically connected by the embedding \emph{in a way that it is possible to obtain them to arbitrary precision for $\kappa$ nonvanishing and finite ($f(x)$ for $\kappa$ sufficiently small and $g(x)$ for $\kappa$ sufficiently large but finite)}. The functions $f(x)$ and $g(x)$ behave as fixed points of the embedding (if we look at the latter dynamically, with $\kappa$ playing the role of a time variable). This strongly contrasts with a linear homotopy in the unit interval
\begin{equation}
h_{q}(x)=qg(x)+(1-q)f(x)=f(x)+q(g(x)-f(x))
\end{equation}
with $q\in \mathbb{R}$, $q\in [0,1]$. We have
\begin{eqnarray}
\lim_{q\to 0}h_{q}(x)&=&f(x) \\
\lim_{q \to 1}h_{q}(x)&=&g(x)
\end{eqnarray}
but
\begin{eqnarray}
\lim_{q \to 0}\frac{\partial h_{q}(x)}{\partial q}=\lim_{q \to 1}\frac{\partial h_{q}(x)}{\partial q}=g(x)-f(x) \ne 0    \end{eqnarray}
so that, at the vicinity of $q \approx 0$ and $q \approx 1$ $h_q(x)$ can be very different to $f(x)$ and $g(x)$ respectively. This contrasts with the nonlinear $\mathcal{B}_{\kappa}$-embeddings here introduced. The latter approach the function $f(x)$ for sufficiently small (but nonvanishing) $\kappa$ and the function $g(x)$ for sufficiently large (but finite) $\kappa$. The parts at zero and infinity are being homotopically connected by sigmoid-like functions. This is apparent in Fig. \ref{embed}. 

The crucial property of the nonlinear embeddings described above make them potentially interesting in the mathematical modeling of complex phenomena which are usually described by e.g. partial differential equations. Because, if one interprets $\kappa$ as time, any arbitrary initial condition can be carried to the same stationary fixed point as time goes to infinity. If there is no fixed point, transient and stationary behaviors can be modeled with nonlinear embeddings as well, although there will also necessarily be a time dependence on the parts carried by the embedding.

The $\mathcal{B}_{\kappa}$-function has remarkable analytic properties in its other arguments as well, and can also be used in the design of continuous differentiable functions with prescribed asymptotic behavior. Let $f(x)$ and $g(x)$ be any continuous and differentiable functions of the real variable $x$. Then, the following function can be constructed that behaves asymptotically like $f(x)$ as $x\to \infty$ and like $g(x)$ as $x\to -\infty$
\begin{equation}
h(x; \kappa)=\frac{f(x)+g(x)}{2}+\frac{f(x)-g(x)}{2}\mathcal{B}_{\kappa}(0,x)
\end{equation}
since, for $\kappa\ge 0$, $\lim_{x\to \infty}\mathcal{B}_{\kappa}(0,x)=-\lim_{x\to \infty}\mathcal{B}_{\kappa}(0,-x)=1$. Furthermore, for $\kappa >0$, the resulting function $h(x; \kappa)$ will be continuous and differentiable if both $f(x)$ and $g(x)$ are.

\section{Compactification in unified physical theories} \label{compact}

We now show how the ideas in the previous section can be applied to model compactification in the passage from supergravity to pure 4-dimensional gravity. The main object of general relativity is the metric tensor with elements $g_{\mu \nu}$ $(\mu, \nu \in \{0,1,2,3\})$ that governs the geometry of spacetime. It is a $4 \times 4$ symmetric matrix ($g_{\mu \nu}=g_{\nu \mu}$) with 10 different components. In our notation, the metric tensor can be specified as
\begin{equation}
g_{\mu \nu}^{(4D)} =\sum_{j=0}^{3}\sum_{k=0}^{3}g_{jk}^{(4D)}\mathcal{B}\left(\mu-j,\frac{1}{2} \right)\mathcal{B}\left(\nu-k,\frac{1}{2} \right)  \qquad (\mu, \nu \in \{0,1,2,3\}) \label{metricten}
\end{equation}

In 11-dimensional supergravity (10-dimensional supergravity would proceed on analogous lines), the metric tensor has the form
\begin{equation}
g_{\mu \nu}^{(11D)} =\sum_{j=0}^{10}\sum_{k=0}^{10}g_{jk}^{(11D)}\mathcal{B}\left(\mu-j,\frac{1}{2} \right)\mathcal{B}\left(\nu-k,\frac{1}{2} \right)  \qquad (\mu, \nu \in \{0,1,2,3,\ldots, 10\}) \label{metrictenM}
\end{equation}
The problem that leads to introduce the idea of compactification is that 11 dimensions are necessary in supergravity (seen as the low energy limit of M-theory) to consistently bring together general relativity and quantum mechanics and, hence, it may be described by a tensor of the form of Eq. (\ref{metrictenM}). However, the observable universe has 4 dimensions and, hence, it is described by an object of the form of Eq. (\ref{metricten}). If we regard the extra 7 dimensions as true, physical dimensions, on a par with the four observed dimensions \cite{Witten81} this suggests to embed both the 4-dimensional metric tensor of pure gravity and the 11-dimensional tensor in a single nonlinear $\mathcal{B}_{\kappa}$-embedding. Let $\ell$ be the scale at which spacetime is probed. If we identify
\begin{equation}
\kappa\propto\frac{\ell}{L_p} \label{PL}
\end{equation}
(where $L_{p}=\sqrt{\frac{G\hbar}{c^3}}=1.6\times 10^{-33}$ cm is the Planck length, with $G$ being the gravitational constant, $c$ the speed of light and $\hbar$ Planck's constant), we can translate the  problem of compactification to the description of a mathematical mechanism involved in the passage from $\mathbf{g}_{11D}$ to $\mathbf{g}_{4D}$ as $\kappa \to \infty$. This suggests the construction of a nonlinear $\mathcal{B}_{\kappa}$-embedding that yields a 4-dimensional metric tensor $\mathbf{g}_{4D}$ in the limit $\kappa \to \infty$ (thus being able to reproduce general relativity in this limit) and the 11-dimensional metric tensor $\mathbf{g}_{11D}$ in the limit $\kappa \to 0$ (thus being potentially useful in unified field theory).

 First, we note that we can rewrite Eq. (\ref{metrictenM}) as
\begin{eqnarray}
g_{\mu \nu}^{(11D)}&=&\sum_{j=0}^{10}\sum_{k=0}^{10}g_{jk}^{(11D)} \mathcal{B}\left(\mu-j,\frac{1}{2} \right)\mathcal{B}\left(\nu-k,\frac{1}{2} \right) \qquad \qquad (\mu, \nu \in \{0,1,2,3,\ldots, 10\}) \label{metrictenM2}\\
&=&\sum_{j=0}^{3}\sum_{k=0}^{3}g_{jk}^{(11D)} \mathcal{B}\left(\mu-j,\frac{1}{2} \right)\mathcal{B}\left(\nu-k,\frac{1}{2} \right)+\sum_{j=4}^{10}\sum_{k=0}^{3}g_{jk}^{(11D)} \mathcal{B}\left(\mu-j,\frac{1}{2} \right)\mathcal{B}\left(\nu-k,\frac{1}{2} \right)\nonumber \\
&&+\sum_{j=0}^{3}\sum_{k=4}^{10}g_{jk}^{(11D)} \mathcal{B}\left(\mu-j,\frac{1}{2} \right)\mathcal{B}\left(\nu-k,\frac{1}{2} \right)+\sum_{j=4}^{10}\sum_{k=4}^{10}g_{jk}^{(11D)}\mathcal{B}\left(\mu-j,\frac{1}{2} \right)\mathcal{B}\left(\nu-k,\frac{1}{2} \right) \nonumber
\end{eqnarray}
We can now apply the replacement Mode I to every $\mathcal{B}$-function involving a dummy index higher or equal than four. In this way, we construct a nonlinear $\mathcal{B}_{\kappa}$-embedding that includes Eq. (\ref{metrictenM}) in the limit $\kappa \to 0$. We obtain
\begin{eqnarray}
g_{\kappa, \mu\nu}&=&\sum_{j=0}^{3}\sum_{k=0}^{3}g_{\kappa, jk}\mathcal{B}\left(\mu-j,\frac{1}{2} \right)\mathcal{B}\left(\nu-k,\frac{1}{2} \right)+\sum_{j=4}^{10}\sum_{k=0}^{3}g_{\kappa,jk}\mathcal{B}_{\kappa}\left(\mu-j,\frac{1}{2} \right)\mathcal{B}\left(\nu-k,\frac{1}{2} \right) \label{THEKEY} \\
&&+\sum_{j=0}^{3}\sum_{k=4}^{10}g_{\kappa,jk}\mathcal{B}\left(\mu-j,\frac{1}{2} \right)\mathcal{B}_{\kappa}\left(\nu-k,\frac{1}{2} \right)+\sum_{j=4}^{10}\sum_{k=4}^{10}g_{\kappa,jk}\mathcal{B}_{\kappa}\left(\mu-j,\frac{1}{2} \right)\mathcal{B}_{\kappa}\left(\nu-k,\frac{1}{2} \right) \nonumber
\end{eqnarray}
We have, by using Eqs. (\ref{lim1}) and (\ref{lim3})
\begin{eqnarray}
\lim_{\kappa \to 0} g_{\kappa, \mu\nu} &=& \sum_{j=0}^{10}\sum_{k=0}^{10}g_{jk}^{(11D)}\mathcal{B}\left(\mu-j,\frac{1}{2} \right)\mathcal{B}\left(\nu-k,\frac{1}{2} \right)=g_{\mu \nu}^{(11D)} \label{Mlim1} \\
\lim_{\kappa \to \infty} g_{\kappa, \mu\nu} &=&  \sum_{j=0}^{3}\sum_{k=0}^{3}g_{jk}^{(4D)}\mathcal{B}\left(\mu-j,\frac{1}{2} \right)\mathcal{B}\left(\nu-k,\frac{1}{2} \right)=g_{\mu \nu}^{(4D)} \label{Mlim2}
\end{eqnarray}
%

If we denote by $(dx^{0}, dx^{1}, \ldots, dx^{11})$ any 11-dimensional spacetime infinitesimal displacement we can write the nonlinear $\mathcal{B}_{\kappa}$-embedding of the squared differential of the arc-length $ds_{\kappa}^2$ as
\begin{eqnarray}
ds_{\kappa}^2&=&\sum_{\mu=0}^{10}\sum_{\nu=0}^{10}g_{\kappa, \mu\nu}dx^{\mu}dx^{\nu} \label{THEds}
\end{eqnarray}
Again, we note that, by using Eqs. (\ref{lim1}) and (\ref{lim3}), we have, consistently 
\begin{eqnarray}
\lim_{\kappa \to 0} ds_{\kappa}^2 &=&\sum_{\mu=0}^{10}\sum_{\nu=0}^{10}g_{\mu\nu}^{(11D)}dx^{\mu}dx^{\nu} \label{MALlim1} \\
\lim_{\kappa \to \infty} ds_{\kappa}^2 &=& \sum_{\mu=0}^{3}\sum_{\nu=0}^{3}g_{\mu\nu}^{(4D)}dx^{\mu}dx^{\nu} \label{MALlim2}
\end{eqnarray}
These equations provide the right differential of the arc-lengths for the respective metric tensors. We thus see that the $\kappa$-deformed structure, Eq. (\ref{THEKEY}) contains the metric tensors of 4-dimensional gravity and 11-dimensional supergravity as specific limiting cases, regardless of the specific form of their elements $g_{\mu\nu}$. Although we have assumed a simple dependence of the parameter $\kappa$ on the Planck length, Eq. (\ref{PL}), this dependence can be more involved and might be derived from first-principles in terms of the local curvature of spacetime, etc. It should be noted that, for $\kappa$ nonvanishing, the $\mathcal{B}_{\kappa}$-function is a smooth and infinitely differentiable function of $\kappa$ and that the nonlinear $\mathcal{B}_{\kappa}$-embeddings given by Eq. (\ref{THEKEY}) and (\ref{THEds}) are smooth, differentiable and well defined for any value of $\kappa$.

\section{Connecting cellular automata and (nonlinear) partial differential equations through nonlinear $\mathcal{B}_{\kappa}$-embeddings} \label{CA2PDE}

In Section \ref{compact} we have shown how nonlinear $\mathcal{B}_{\kappa}$-embeddings can be used to construct generalized structures that connect mathematical objects with different dimensionality. In this and the following section, we show how they can be used indeed to smoothly connect different qualitative dynamical behaviors governed by different dynamical (evolution) rules. An increased value of the continuous parameter $\kappa \in \mathbb{R}$ is in all the following associated to the loosening of the rigidity of the dynamical rules (CA-like rules) found in the limit $\kappa \to 0$. Consequently, an increased value of the reciprocal $1/\kappa$ represents the loosening of the dynamical rules found at $\kappa \to \infty$. Thus, an increased value of $\kappa$ provides a smooth, continuous and differentiable connection between two limits in which dynamical rules can be either discrete or continuous. If the dynamics is continuous, the embeddings converge uniformly to the limiting cases. If it is discrete, convergence is, necessarily, pointwise. Note that, depending on the physical application, $\kappa$ can be regarded as a temperature-like parameter (so that when $\kappa$ is increased a 'thermal motion' is intensified that weakens the dynamical rules) or even as time. Nonlinear $\mathcal{B}_{\kappa}$-embeddings can therefore be used to construct `cartographies' of physical theories and bifurcation scenarios where qualitative changes in dynamical behavior are induced by tuning the continuous parameter $\kappa$. In these cartographies, theories (specific models) are encompassed by more advanced theories in a hierarchical manner.

CAs \cite{Wolfram, DeutschBOOK, Tokihiro1, Ilachinski, Adamatzky, Wuensche, VGM1, semipredo,VGM2,VGM3,diagrammar}, CMLs \cite{Kaneko1, Kaneko2, Kapral, KanekoBOOK, Bunimovich2, Bunimovich3,JPHYSA} and (nonlinear) PDEs \cite{Hohenberg, KuramotoBOOK} constitute the different mathematical approaches to model spatiotemporal pattern formation outside of equilibrium, as found in experimental physical systems \cite{DeutschBOOK}. CAs are fully discrete coupled maps in which space and time are discrete and the local phase space is both discrete and finite. CAs serve as toy models for the overall observed features of complex physical systems \cite{Wolfram}. An example of this is spatiotemporal intermittency \cite{Wolfram, Jabeen1, Jabeen2}. CMLs \cite{Kaneko1, Kaneko2} are discrete maps ruling the evolution of a dynamical system on a discrete spacetime but for which the local phase space is continuous. Finally, PDEs constitute continuous models of dynamical system evolving on continuous and differentiable spacetimes and with a continuous local phase space. In this section we show how all CAs can be encompassed by means of $\mathcal{B}_{\kappa}$-embeddings that connect them to certain CMLs and nonlinear PDEs. In particular, we show how some CAs lead to nonlinear diffusion equations.

We define the alphabet $\mathcal{A}_{p}\equiv \{0,1,\ldots,p-1\}$, $p\ge 2$, $p\in \mathbb{N}$, as the set of integers in the interval $[0,p-1]$. We write $\mathcal{A}_{p}^{N}$ for the Cartesian product of $N$ copies of $\mathcal{A}_{p}$.  Let $x_{t}^{j}\in \mathcal{A}_{p}$ be a dynamical variable at time $t\in \mathbb{Z}$ and position $j\in \mathbb{Z}$ on a ring of $N_{s}$ sites, $j \in [0,N_{s}-1]$ and let $l$, $r$ be non-negative integers. A CA, with rule vector $(a_{0},a_{1},\ldots, a_{p^{l+r+1}-1})$, $a_{n\in \mathbb{Z}}\in \mathcal{A}_{p},\ \forall n\in[0,p^{l+r+1}-1]$, range $N=l+r+1$ and Wolfram code $R=\sum_{n=0}^{p^{l+r+1}-1}a_{n}p^{n}$, is a map $\mathcal{A}_{p}^{N_s}\to \mathcal{A}_{p}^{N_s}$ acting locally at each site $j$ as $\mathcal{A}_{p}^{N}\to \mathcal{A}_{p}$ and synchronously at every $t$ according to the universal map \cite{VGM1}
\begin{equation}
x_{t+1}^{j}=\sum_{n=0}^{p^{r+l+1}-1}a_{n}\mathcal{B}\left(n-\sum_{k=-r}^{l}p^{k+r}x_{t}^{j+k},\frac{1}{2}\right) \qquad j=0,1,\ldots, N_{s}-1 \label{CA}
\end{equation} 
where $j+k=j+k \mod N_{s}$. We note that the Wolfram code $R$ is an integer $R\in [0,p^{p^{l+r+1}}]$.

All parameters specifying any CA rule can be given in a compact notation by means of the code $^{l}R^{r}_{p}$ \cite{VGM1}. For example, all 256 Wolfram elementary CA $^{1}R^{1}_{2}$ are obtained by taking $p=2$, $l=r=1$ in Eq. (\ref{CA}). Thus, Wolfram rule 30 is denoted by $^{1}30^{1}_{2}$ and has rule vector $(a_{0},a_{1},\ldots a_{7})=(0,1,1,1,1,0,0,0)$. In general, the coefficients $a_{n}$ of any CA can be directly obtained from the Wolfram code $R$ by means of the following expression \cite{semipredo}   
\begin{equation}
a_{n}= \left \lfloor \frac{R}{p^{n}} \right \rfloor-p\left \lfloor \frac{R}{p^{n+1}} \right \rfloor    \label{cucuAreal}
\end{equation}
where $\left \lfloor \ldots \right \rfloor$ denotes the lower closest integer (floor) function. 

Specially interesting for physical applications are those CAs that are locally isotropic so that the CA output does not depend on the particular arrangement of the dynamical states within a neighborhood, but on the sum of the cell values. These are called totalistic CAs and are a subset of those described by Eq. (\ref{CA}). Totalistic CAs are given by the map
\begin{equation}
x_{t+1}^{j}=\sum_{n=0}^{(r+l+1)(p-1)}\sigma_{n}\mathcal{B}\left(n-\sum_{k=-r}^{l}x_{t}^{j+k},\frac{1}{2}\right) \qquad j=0,1,\ldots, N_{s}-1 \label{CAT}
\end{equation} 
where $\sigma_{n}\in \mathcal{A}_{p}$. The Wolfram code of a totalistic CA is given by $RT\equiv \sum_{n=0}^{(r+l+1)(p-1)}\sigma_{n}p^{n}$ and is an integer number satisfying $RT\in [0,p^{1+(r+l+1)(p-1)}]$.

Let the real line excluding all half-integer numbers ($\ldots, -\frac{3}{2}, -\frac{1}{2}, \frac{1}{2}, \frac{3}{2},\ldots$) be denoted by $\mathbb{R}_{\setminus \mathbb{Z}/2}$. From the definition of the $\mathcal{B}$-function, it is clear that the function $\mathcal{B}\left(n-x,\frac{1}{2} \right)$ for $n\in \mathbb{Z}$ and $x\in \mathbb{R}_{\setminus \mathbb{Z}/2}$ provides a surjective application $\mathbb{Z}\times\mathbb{R}_{\setminus \mathbb{Z}/2} \to \{0,1\}$. Thus, if we relax any $x_t^j$ in Eq. (\ref{CA}) to be a real number so that $n_t^j\equiv \sum_{k=-r}^{l}p^{k+r}x_{t}^{j+k}\in  \mathbb{R}_{\setminus \mathbb{Z}/2}$, we find that $x_{t+1}^j\in \mathcal{A}_{p}$. Therefore, for all initial conditions $x_0^j$ for which $n_0^j= \sum_{k=-r}^{l}p^{k+r}x_{0}^{j+k} \in \mathbb{R}_{\setminus \mathbb{Z}/2}$ ($\forall j$) the local dynamics provided by Eq. (\ref{CA}) has the form $\mathcal{A}_{p}^{N}\to \mathcal{A}_{p}$ for any $t>0$. This fact allowed us to generalize in \cite{JPHYSA}  the universal map for CA, Eq. (\ref{CA}), to real-valued deterministic CA in terms of the $\mathcal{B}_{\kappa}$-function, Eq. (\ref{bkappa})
\begin{equation}
x_{t+1}^{j}=\sum_{n=0}^{p^{r+l+1}-1}a_{n}\mathcal{B}_{\kappa}\left(n-\sum_{k=-r}^{l}p^{k+r}x_{t}^{j+k},\frac{1}{2}\right) \qquad j=0,1,\ldots, N_{s}-1 \label{RDCA}
\end{equation} 
This generalization amounts to use in Eq. (\ref{CA}) the replacement Mode I described in Section \ref{embeddings}. Therefore, by using Eq. (\ref{lim2}), Eq. (\ref{RDCA}) becomes equal to Eq. (\ref{CA}) in the limit $\kappa \to 0$. In the limit $\kappa \to \infty$ one has, from Eq. (\ref{Bernoultheo1}), $x_{t+1}^{j} \sim \frac{1}{2\kappa}$ \cite{JPHYSA} so that, if the limit is strictly taken $x_{t}^{j}=0, \forall j$ and $\forall t>0$. 

If we use Mode II instead, we obtain
\begin{equation}
x_{t+1}^{j}=\sum_{n=0}^{p^{r+l+1}-1}a_{n}\frac{\mathcal{B}_{\kappa}\left(n-\sum_{k=-r}^{l}p^{k+r}x_{t}^{j+k},\frac{1}{2}\right)}{\mathcal{B}_{\kappa}\left(0,\ \frac{1}{2}\right)} \qquad j=0,1,\ldots, N_{s}-1 \label{RDCA2}
\end{equation} 
the limit $\kappa \to 0$ is as before but now the limit $\kappa \to \infty$ is
\begin{equation}
x_{t+1}^{j}=\sum_{n=0}^{p^{r+l+1}-1}a_{n} \qquad j=0,1,\ldots, N_{s}-1 \label{RDCA2lim}
\end{equation} 
because of Eq. (\ref{lim1}). Eqs. (\ref{RDCA}) and (\ref{RDCA2}) describe respective CMLs taking place on the real numbers. We thus see how $\mathcal{B}_{\kappa}$-embeddings connect CA and certain CMLs.

If we consider the replacement Mode II on totalistic CA, we have, from Eqs. (\ref{CAT}) and (\ref{Bernoultheo2})
\begin{eqnarray}
x_{t+1}^{j}&=&\sum_{n=0}^{(r+l+1)(p-1)}\sigma_{n}\frac{\mathcal{B}_{\kappa}\left(n-\sum_{k=-r}^{l}x_{t}^{j+k},\frac{1}{2}\right)}{\mathcal{B}_{\kappa}\left(0,\frac{1}{2}\right)} \qquad j=0,1,\ldots, N_{s}-1 \\ 
&=&\sum_{n=0}^{(r+l+1)(p-1)}\sigma_{n}\left[1-\frac{\left(n-\sum_{k=-r}^{l}x_{t}^{j+k}\right)^2}{\kappa^2}+O\left(\kappa^{-4}\right) \right] \qquad j=0,1,\ldots, N_{s}-1
\end{eqnarray}
where the last equation is obtained in the asymptotic limit $\kappa$ large. Let us consider, more specifically, local rules for which $l=r=1$. We obtain
\begin{eqnarray}
x_{t+1}^{j}&=&\sum_{n=0}^{3(p-1)}\sigma_{n}\left[1-\frac{\left(n-x_{t}^{j+1}-x_{t}^{j}-x_{t}^{j-1}\right)^2}{\kappa^2}+O\left(\kappa^{-4}\right) \right] \qquad j=0,1,\ldots, N_{s}-1 \nonumber \\
&=&\sum_{n=0}^{3(p-1)}\sigma_{n}\left[1-\frac{\left(n-x_{t}^{j+1}+2x_{t}^{j}-x_{t}^{j-1}-3x_{t}^{j}\right)^2}{\kappa^2}+O\left(\kappa^{-4}\right) \right] \qquad j=0,1,\ldots, N_{s}-1 \nonumber \\
&=& x_{t}^{j}+P\left(x_{t}^{j}\right)+D\Delta^2_{L} x_{t}^{j}-F\left(x_{t}^{j}, \Delta^2_{L} x_{t}^{j}\right)
\end{eqnarray}
where we have defined
\begin{eqnarray}
\Delta_{L}^2 x_{t}^{j} &\equiv & x_{t}^{j+1}-2x_{t}^{j}+x_{t}^{j-1} \label{secondiff}\\
D&\equiv& \frac{2}{\kappa^2}\sum_{n=0}^{3(p-1)}\sigma_{n}n \label{Ddef} \\
P\left(x_{t}^{j}\right)&\equiv &-x_{t}^{j}+\sum_{n=0}^{3(p-1)}\sigma_{n}\left[1-\frac{\left(n-3x_{t}^{j}\right)^2}{\kappa^2} \right] \label{Pdef}\\
F\left(x_{t}^{j}, \Delta^2 x_{t}^{j}\right) &\equiv& \left(6x_{t}^j\Delta^2 x_{t}^{j} + (\Delta^2 x_{t}^{j} )^2   \right) \frac{\sum_{n=0}^{3(p-1)}\sigma_{n}}{\kappa^2} +O\left(\kappa^{-4}\right)  \label{Ndef} 
\end{eqnarray}
By further introducing 
\begin{equation}
\Delta_{T} x_{t}^{j} \equiv x_{t+1}^{j}-x_{t}^{j} \label{firstdiff}
\end{equation}
we, finally, obtain
\begin{equation}
\Delta_{T} x_{t}^{j}= P\left(x_{t}^{j}\right)+D\Delta^2_{L} x_{t}^{j}-F\left(x_{t}^{j}, \Delta^2_{L} x_{t}^{j}\right) \label{discredifu}
\end{equation} 
This is a difference equation involving a discretized Laplacian $\Delta^2_{L} x_{t}^{j}$ and the first-order time difference $\Delta_{T} x_{t}^{j}$. By making the following transformations
\begin{eqnarray}
t&\to &N_{T}\tau \\
t+1&\to &(N_{T}+1)\tau \\
j&\to &N_{L}\ell \\
j+1&\to &(N_{L}+1)\ell \\
j-1&\to &(N_{L}-1)\ell \\
x &\equiv& j=N_{L}\ell \\
u(t,x)&\equiv& x_{t}^{j} 
\end{eqnarray}
and by taking the limits $N_{T} \to \infty$, $N_{L} \to \infty$, $\tau \to 0$, $\ell \to 0$, so that $N_{T}\tau=t$, $N_{L}\ell=x$ remain constant, we obtain from Eq. (\ref{discredifu})
\begin{equation}
\tau\frac{\partial u}{\partial t}= P\left(u\right)+\ell^2 D\frac{\partial^2 u}{\partial x^2}-F\left(u, \ell^2\frac{\partial^2 u}{\partial x^2}\right) \label{difunolin}
\end{equation}
The function $P\left(x_{t}^{j}\right)$ is a quadratic polynomial governing the homogeneous dynamics. The function $F\left(x_{t}^{j}, \Delta^2_{L} x_{t}^{j}\right)$ is nonlinearly dependent on the discretized Laplacian and the cell dynamical state $x_{t}^{j}$ and can be thought as a kind of `nonlinear diffusion' term. If this last term vanishes, Eq. (\ref{difunolin}) is a Fisher-Kolmogorov equation.
Furthermore, a linear diffusion equation is always obtained in the limit $\kappa$ large if the following conditions are met
\begin{eqnarray}
\sum_{n=0}^{3(p-1)}\sigma_{n} &=& 0 \label{condium1}\\
\sum_{n=0}^{3(p-1)}n\sigma_{n} &>& 0 \label{condium2}
\end{eqnarray}
in which case Eq. (\ref{difunolin}) simplifies to
\begin{equation}
\frac{\partial u}{\partial t}= f\left(u\right)+D\frac{\partial^2 u}{\partial x^2} \label{difulin}
\end{equation}
where
\begin{eqnarray}
f(u)&=&(3D-1)\frac{u}{\tau}-\frac{1}{\tau\kappa^2}\sum_{n=0}^{3(p-1)}n^2\sigma_{n} \label{fudef}\\
D&\equiv& \frac{2\ell^2}{\tau\kappa^2}\sum_{n=0}^{3(p-1)}n\sigma_{n} \label{Ddef2} 
\end{eqnarray}
If $3D<1$ (which can always be the case for $\kappa$ sufficiently large) the homogeneous dynamics of Eq. (\ref{difulin}) converges to the stable homogeneous fixed point given by $f(u^{*})=0$, i.e.
\begin{equation}
u^{*}=\frac{1}{(3D-1)\kappa^2}\sum_{n=0}^{3(p-1)}n^2\sigma_{n}
\end{equation}
We note that the conditions expressed by Eqs. (\ref{condium1}) and (\ref{condium2}) are curiously the same as those found in the construction of appropriate difference operators for generalized logarithms and group entropies (see Eq. (2) in \cite{Tempesta}).

\section{Cellular automata connections} \label{CMLcon}

We now show how any two CAs in rule space can be connected by means of a nonlinear $\mathcal{B}_{\kappa}$-embedding so that in the limits $\kappa \to 0$ and $\kappa \to \infty$ each of the CAs entering in the connection is obtained.  We call such nonlinear $\mathcal{B}_{\kappa}$-embedding, generally behaving as a CML, a CA connection. The embedded CAs are called the CA limits of the CA connection. 

For intermediate $\kappa$ values, we derive a  mean-field model of the connection that can qualitatively capture many of its dynamical features, as observed in its spatiotemporal evolution. In general, a finite non-vanishing value of the parameter $\kappa$ weakens the `pure' behavior of the CA limits and we believe that the general concept of CA connections introduced here may be useful in, e.g. biophysical models of multicellular ensembles \cite{Hwang}, where variability and network heterogeneity need to be taken into account and may incorporate several kinds of typical CA dynamics. In these applications $\kappa$ may be related to biological time/or and to the connectivity of the multicellular ensemble mediated by gap junctions \cite{VGMSMJA}. The strength of the coupling is given by $1/\kappa$ so that, when $\kappa \to 0$ (strong coupling regime), the cells in the ensemble are tightly coupled and one can qualitatively describe the ensemble by means of a CA. However, aging of the network may lead to a lower value of $1/\kappa$ leading to conformational changes that, in turn, may lead to dynamical changes so that the network is loosened. Finally, the presence of other agents in the network, facilitated by the decreased network connectivity may induce a different CA dynamics that is qualitatively different than the one obtained in the limit $\kappa \to 0$. In this article we construct the whole general class of CA connections (CMLs) in which any two arbitrary CAs in rule space can be present in the CA limits. 

The question now arises whether we can modify Eq. (\ref{RDCA2}) so that  the limit $\kappa \to \infty$ is another cellular automaton of the form of Eq. (\ref{CA}) for all initial conditions $x_0^j \in \mathcal{A}_p$ ($\forall j$) but with a generally different rule vector $(b_{0},b_{1},\ldots, b_{p^{l+r+1}-1})$, $b_{n\in \mathbb{Z}}\in \mathcal{A}_{p},\ \forall n\in[0,p^{l+r+1}-1]$
\begin{equation}
x_{t+1}^{j}=\sum_{n=0}^{p^{r+l+1}-1}b_{n}\mathcal{B}\left(n-\sum_{k=-r}^{l}p^{k+r}x_{t}^{j+k},\frac{1}{2}\right) \qquad j=0,1,\ldots, N_{s}-1 \label{CA2}
\end{equation}

We construct a $\kappa$-deformed formula so that two such CA are connected in the limits $\kappa \to 0$ and $\kappa \to \infty$. We note that $x_{t+1}^{j} \in \mathcal{A}_{p}$ is any of the integers $m$ in the interval $[0,p-1]$. Therefore, we observe that
\begin{eqnarray}
x_{t+1}^{j}&=&\sum_{m=0}^{p-1}m\delta_{m x_{t+1}^{j}} =\sum_{m=0}^{p-1}m\mathcal{B}\left(m-x_{t+1}^{j}\right) \nonumber \\
&=&\sum_{m=0}^{p-1}m\mathcal{B}\left(m-x_{t+1}^{j}\right)\mathcal{B}\left(m-x_{t+1}^{j}\right)
\end{eqnarray}
By using the replacement Mode II and the $\kappa \to 1/\kappa$ transformation, we construct from here a nonlinear $\mathcal{B}_{\kappa}$-embedding
\begin{equation}
f_{\kappa}(x_{t+1}^{j})=\sum_{m=0}^{p-1}m\frac{\mathcal{B}_{\kappa}\left(m-x_{t+1}^{j} ,\frac{1}{2}\right)}{\mathcal{B}_{\kappa}(0,\frac{1}{2})}\frac{\mathcal{B}_{1/\kappa}\left(m-x_{t+1}^{j} ,\frac{1}{2}\right)}{\mathcal{B}_{1/\kappa}(0,\frac{1}{2})}\qquad j=0,1,\ldots,N_s-1 \label{CAconPRE}
\end{equation}
By defining,
\begin{eqnarray}
a_{t}^{j}(\kappa)&\equiv&\sum_{n=0}^{p^{l+r+1}-1}a_{n}\mathcal{B}_{\kappa}\left(n-\sum_{k=-r}^{l}p^{k+r}x_{t}^{j+k},\frac{1}{2}\right) \label{finfi}\\
b_{t}^{j}(\kappa)&\equiv&\sum_{n=0}^{p^{l+r+1}-1}b_{n}\mathcal{B}_{1/\kappa}\left(n-\sum_{k=-r}^{l}p^{k+r}x_{t}^{j+k},\frac{1}{2}\right) \label{g0}
\end{eqnarray}
we can construct from Eq. (\ref{CAconPRE}) a CA connection involving two CAs $^{l}[R_{1}]_{p}^{r}$ and $^{l}[R_{2}]_{p}^{r}$ in the CA limits of the connection   
\begin{equation}
x_{t+1}^{j}=\sum_{m=0}^{p-1}m\frac{\mathcal{B}_{\kappa}\left(m-a_{t}^{j}(\kappa) ,\frac{1}{2}\right)}{\mathcal{B}_{\kappa}(0,\frac{1}{2})}\frac{\mathcal{B}_{1/\kappa}\left(m-b_{t}^{j}(\kappa) ,\frac{1}{2}\right)}{\mathcal{B}_{1/\kappa}(0,\frac{1}{2})}\qquad j=0,1,\ldots,N_s-1 \label{CAcon}
\end{equation}
We denote a CA connection by  $^{l}[R_{1}]_{p}^{r}\ \xrightarrow{\kappa}\ ^{l}[R_{2}]_{p}^{r}$. In the limit $\kappa\to 0$, Eq. (\ref{CAcon}) becomes Eq. (\ref{CA}) since
\begin{eqnarray}
\lim_{\kappa\to 0}x_{t+1}^{j}&=&\lim_{\kappa\to 0}\sum_{m=0}^{p-1}m\frac{\mathcal{B}_{\kappa}\left(m-a_{t}^{j}(\kappa) ,\frac{1}{2}\right)}{\mathcal{B}_{\kappa}(0,\frac{1}{2})}\frac{\mathcal{B}_{1/\kappa}\left(m-b_{t}^{j}(\kappa) ,\frac{1}{2}\right)}{\mathcal{B}_{1/\kappa}(0,\frac{1}{2})} \nonumber \\
&=&\sum_{m=0}^{p-1}m\mathcal{B}\left(m-a_{t}^{j}(0) ,\frac{1}{2}\right)=a_{t}^{j}(0)=\sum_{n=0}^{p^{l+r+1}-1}a_{n}\mathcal{B}\left(n-\sum_{k=-r}^{l}p^{k+r}x_{t}^{j+k},\frac{1}{2}\right)
\end{eqnarray}
However, in the limit $\kappa\to \infty$ Eq. (\ref{CAcon}) becomes Eq. (\ref{CA2})
\begin{eqnarray}
\lim_{\kappa\to \infty}x_{t+1}^{j}&=&\lim_{\kappa\to \infty}\sum_{m=0}^{p-1}m\frac{\mathcal{B}_{\kappa}\left(m-a_{t}^{j}(\kappa) ,\frac{1}{2}\right)}{\mathcal{B}_{\kappa}(0,\frac{1}{2})}\frac{\mathcal{B}_{1/\kappa}\left(m-b_{t}^{j}(\kappa) ,\frac{1}{2}\right)}{\mathcal{B}_{1/\kappa}(0,\frac{1}{2})} \nonumber \\
&=&\sum_{m=0}^{p-1}m\mathcal{B}\left(m-b_{t}^{j}(\infty) ,\frac{1}{2}\right)=b_{t}^{j}(\infty)=\sum_{n=0}^{p^{l+r+1}-1}b_{n}\mathcal{B}\left(n-\sum_{k=-r}^{l}p^{k+r}x_{t}^{j+k},\frac{1}{2}\right)
\end{eqnarray}
We note that the transformation $\kappa \to 1/\kappa$ merely reverses the connection, i.e. $^{l}[R_{1}]_{p}^{r}\ \xrightarrow{\kappa}\ ^{l}[R_{2}]_{p}^{r}$ changes to $^{l}[R_{2}]_{p}^{r}\ \xrightarrow{\kappa}\ ^{l}[R_{1}]_{p}^{r}$.

For Boolean CA connections ($p=2$) Eq. (\ref{CAcon}) simplifies to
\begin{equation}
x_{t+1}^{j}=\frac{\mathcal{B}_{\kappa}\left(1-a_{t}^{j}(\kappa) ,\frac{1}{2}\right)}{\mathcal{B}_{\kappa}(0,\frac{1}{2})}\frac{\mathcal{B}_{1/\kappa}\left(1-b_{t}^{j}(\kappa) ,\frac{1}{2}\right)}{\mathcal{B}_{1/\kappa}(0,\frac{1}{2})}\qquad j=0,1,\ldots,N_s-1 \label{CAconB}
\end{equation}
and, by the definition of the $\mathcal{B}_\kappa$ function we find that $0\le x_{t}^{j} \le 1$ for all $\kappa$, $t$ and $j$. 

To get insight in the complex spatiotemporal dynamics of Eq. (\ref{CAcon}) it proves useful to consider the mean field model obtained by taking $x_t^j\equiv u_t$. Then Eq. (\ref{CAcon}) reduces to
\begin{equation}
u_{t+1}=\sum_{m=0}^{p-1}m\frac{\mathcal{B}_{\kappa}\left(m-a_{t}(\kappa) ,\frac{1}{2}\right)}{\mathcal{B}_{\kappa}(0,\frac{1}{2})}\frac{\mathcal{B}_{1/\kappa}\left(m-b_{t}(\kappa) ,\frac{1}{2}\right)}{\mathcal{B}_{1/\kappa}(0,\frac{1}{2})} \label{CAconMF}
\end{equation}
where
\begin{eqnarray}
a_{t}(\kappa)&\equiv&\sum_{n=0}^{p^{l+r+1}-1}a_{n}\mathcal{B}_{\kappa}\left(n-\frac{p^{l+r+1}-1}{p-1}u_t,\frac{1}{2}\right) \label{finfiMF}\\
b_{t}(\kappa)&\equiv&\sum_{n=0}^{p^{l+r+1}-1}b_{n}\mathcal{B}_{1/\kappa}\left(n-\frac{p^{l+r+1}-1}{p-1}u_t,\frac{1}{2}\right) \label{g0MF}
\end{eqnarray} 
This reduced model describes the behavior of homogeneous initial conditions and is also a mean field approximation for arbitrary (inhomogeneous) initial conditions. Of special interest are the $\omega$-limit sets of Eq. (\ref{CAconMF}) and their change with the control parameter $\kappa$. These yield the bifurcation diagram of the reduced model that can be used to interpret certain results obtained with the full model. The $\omega$-limit sets of CMLs can be used to model the vacuum fluctuations of chaotic strings (which are specific one-dimensional CMLs underlying the Parisi-Wu approach of stochastic quantization on a small scale) \cite{Beck2}.

\begin{figure}
\includegraphics[width=0.8 \textwidth]{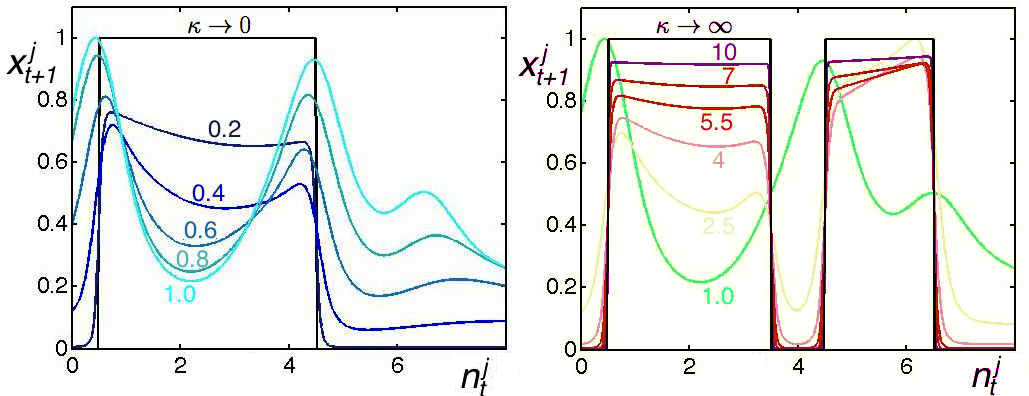}
\caption{\scriptsize{The cell state $x_{t+1}^{j}\in [0,1]$ vs. the neighborhood sum value $n_{t}^{j}=\sum_{k=-1}^{1}2^{k+r}x_{t}^{j+k}$ for the CA connection $^{1}30_{2}^{1}\ \xrightarrow{\kappa}\ ^{1}110_{2}^{1}$ and for the values of $\kappa$ indicated on the curves. The panels are separated to better show the changes obtained in Eq. (\ref{CAconB})  with increasing $\kappa$ values.}} \label{1}
\end{figure}

As an example to illustrate the above concepts, we consider a CA connection between the two elementary CA provided by the Boolean CA rules $^{1}30^{1}_2$ and $^{1}110^{1}_2$. The former is known to be a random number generator and the latter is capable of universal computation \cite{Wolfram}. We thus consider the CA connection $^{1}30_{2}^{1}\ \xrightarrow{\kappa}\ ^{1}110_{2}^{1}$ described by Eq. (\ref{CAconB}) with $p=2$, $l=r=1$ and rule vectors $(a_0,a_1,\ldots,a_7)=(0,1,1,1,1,0,0,0)$ and $(b_0,b_1,\ldots,b_7)=(0,1,1,1,0,1,1,0)$.  
Fig. \ref{1} shows $x_{t+1}^{j}$ in Eq. (\ref{CAconB}) as a function of the neighborhood value $n_{t}^{j}=\sum_{k=-1}^{1}p^{k+r}x_{t}^{j+k}$ for different values of $\kappa$ and for the above CA connection: Note that CA rule $^{1}30_{2}^{1}$ is obtained in the limit $\kappa \to 0$ (Fig. \ref{1} left) while rule $^{1}110_{2}^{1}$ is obtained in the limit $\kappa \to \infty$ (Fig. \ref{1} right). For intermediate values of $\kappa$ the curves always lie within the unit interval.

\begin{figure}
\includegraphics[width=1.0 \textwidth]{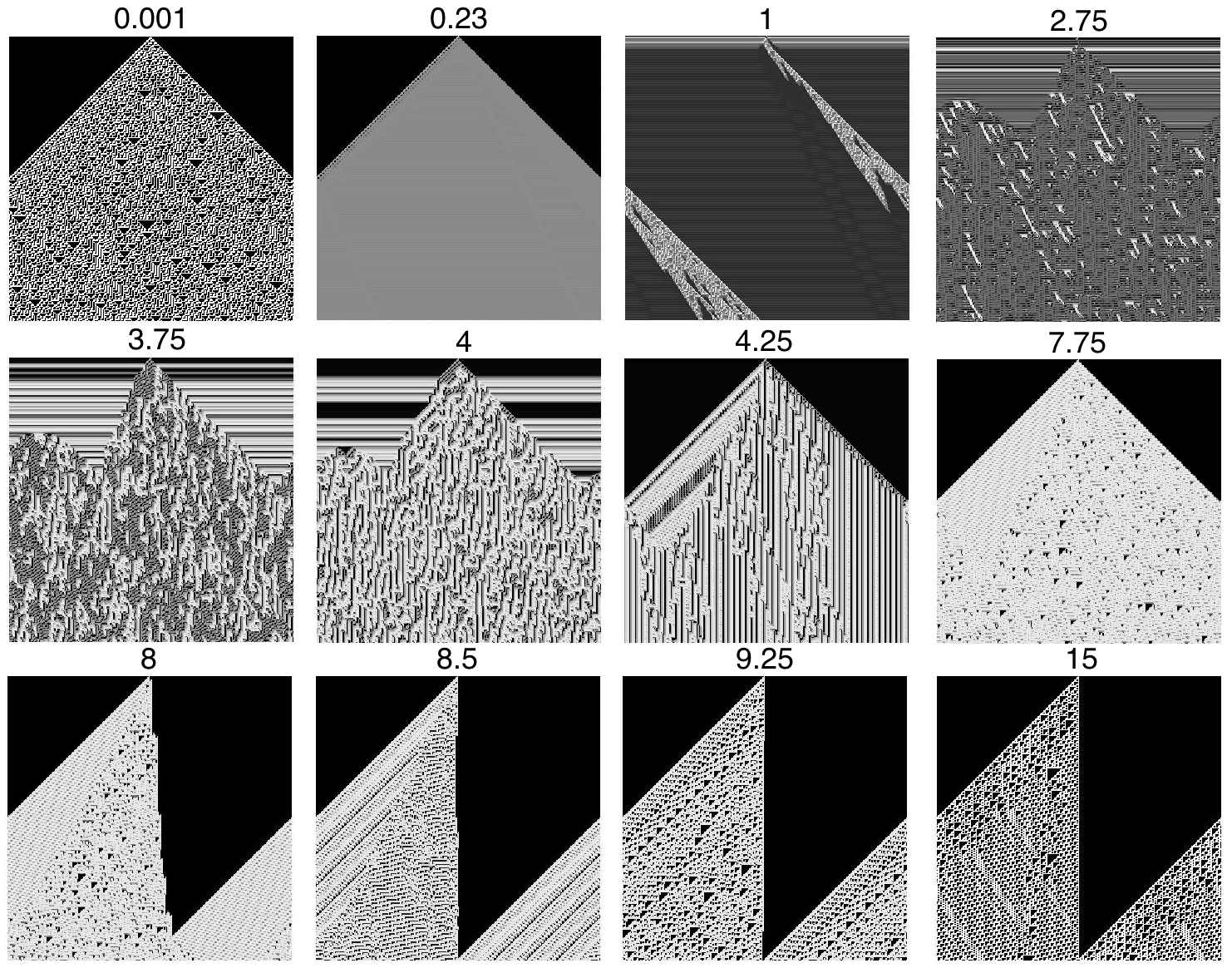}
\caption{\scriptsize{Spatiotemporal evolution of $x_t^j$ obtained from Eq. (\ref{CAconB}) for the CA connection $^{1}30_{2}^{1}\ \xrightarrow{\kappa}\ ^{1}110_{2}^{1}$ starting from a single site with value 1 surrounded by sites with zero value and for the values of $\kappa$ indicated on the panels. In each panel, time flows from top to bottom. The system size is $N_s=240$ and 240 iteration steps are shown. In the gray scale used, black and white correspond to values zero and one, respectively.}} \label{2}
\end{figure}

The spatiotemporal evolution of $x_t^j$ obtained from Eq. (\ref{CAconB}) for the CA connection $^{1}30_{2}^{1}\ \xrightarrow{\kappa}\ ^{1}110_{2}^{1}$ is shown in Fig. \ref{2} $j\in [0,239]$, $t\in [0,239]$ and for different $\kappa$ values indicated over the panels for a simple initial condition consisting of a single site with value '1' surrounded by sites with value '0'. For $\kappa=0.001$ the behavior typical of the limit $\kappa \to 0$ is already observed and the CA connection coincides with Wolfram's CA rule 30. For $\kappa=15$ the limit $\kappa \to \infty$ is already well approximated and the CA connection coincides with Wolfram's CA rule 110. For intermediate values of $\kappa$ the behavior is highly nontrivial and gliders and coherent structures typical of Class 4 CA \cite{Wolfram, Ilachinski} are observed (e.g. for $\kappa=4.25$). The overall behavior for $\kappa$ finite somehow interpolates between the limits $\kappa \to 0$ and $\kappa \to \infty$, although there is a wide variety of qualitatively different behavior. At $\kappa=1$ a coexistence is observed between a traveling turbulent patch at high values of $x_t^j$ and homogeneous period-2 oscillations that form domains (clusters) connecting two different low values of $x_t^j$. Most remarkably, it is observed that in the interval $2.75\lesssim \kappa \lesssim 4$, the homogeneous quiescent state loses stability to aperiodic oscillations, the system being highly sensitive to small perturbations.

Insight on the results of Fig. \ref{2} can be gained by means of the mean-field reduction of the CA connection, Eq. (\ref{CAconMF}) which, in this particular case, takes the form
\begin{equation}
u_{t+1}=\frac{\mathcal{B}_{\kappa}\left(1-a_{t}(\kappa) ,\frac{1}{2}\right)}{\mathcal{B}_{\kappa}(0,\frac{1}{2})}\frac{\mathcal{B}_{1/\kappa}\left(1-b_{t}(\kappa) ,\frac{1}{2}\right)}{\mathcal{B}_{1/\kappa}(0,\frac{1}{2})} \label{CAconMFB}
\end{equation}
where
\begin{eqnarray}
a_{t}(\kappa)&\equiv&\sum_{n=0}^{7}a_{n}\mathcal{B}_{\kappa}\left(n-7u_t,\frac{1}{2}\right) \label{finfiMFB}\\
b_{t}(\kappa)&\equiv&\sum_{n=0}^{7}b_{n}\mathcal{B}_{1/\kappa}\left(n-7u_t,\frac{1}{2}\right) \label{g0MFB}
\end{eqnarray} 
with $(a_0,a_1,\ldots,a_7)=(0,1,1,1,1,0,0,0)$ and $(b_0,b_1,\ldots,b_7)=(0,1,1,1,0,1,1,0)$.   

\begin{figure}
\includegraphics[width=0.7 \textwidth]{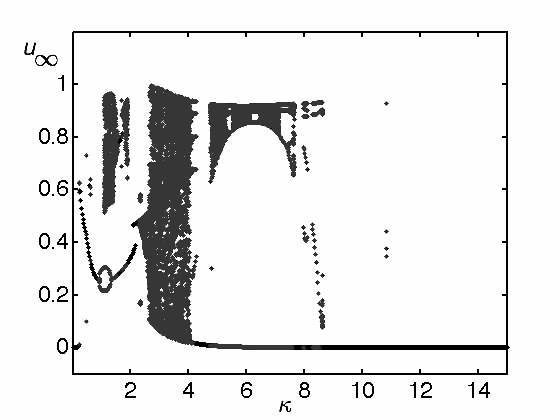}
\caption{\scriptsize{Bifurcation diagram calculated from the asymptotic behavior of Eq. (\ref{CAconMFB}) for the CA connection $^{1}30_{2}^{1}\ \xrightarrow{\kappa}\ ^{1}110_{2}^{1}$. The values that the dynamical variable $u_{\infty}$ takes on the orbit at large times are shown as a function of the parameter $\kappa$.}} \label{3}
\end{figure}

The bifurcation diagram can be readily calculated by iterating Eq. (\ref{CAconMFB}) for sufficiently long times, starting from initial conditions $u_{0}\in [0,1]$ uniformly filling the unit interval. In this way, the $\omega$-limit sets $u_{\infty}$ of the values of $u_t$ as $t\to \infty$ are numerically obtained. The bifurcation diagram is shown in Fig. \ref{3} in which $u_{\infty}$ is plotted vs. $\kappa$.  

Some observations made in Fig. \ref{2} can be qualitatively understood by means of the bifurcation diagram in Fig. \ref{3}:
 \begin{itemize}
\item In the limits $\kappa \to 0$ and $\kappa \to \infty$ we have $u_{\infty}\to 0$. This corresponds to the CA limits and both rules $^{1}30^{1}_{2}$ and $^{1}110^{1}_{2}$ in the connection fix the quiescent state and no other homogeneous configuration. The mean-field model is not able to capture the complex dynamics of Eq. (\ref{CAconB}) in these CA limits because the dynamical behavior is highly correlated through the CA dynamics and the correlations, that are affected by the neighborhood configurations, are lost in the mean-field model. 
\item For $\kappa \approx 1$ the coexistence between two alternating period-2 branches at low values of $u_{\infty}$ and a chaotic stripe at high values of $u_{\infty}$ is observed. A gap in $u_{\infty}$ is seen separating both behaviors. This qualitatively captures the observation made in Fig. \ref{2} for this value of $\kappa$.
\item In the interval $2.75\lesssim \kappa \lesssim 4$, $u_\infty$ displays a wide variety of possible states that are all reached in an aperiodic, chaotic, manner. The branch at low $u_\infty$ is now fused with the chaotic stripe and there is no gap. This matches the observation made on the turbulent behavior described in Fig. \ref{2} and the mean-field model allows to relate that turbulence in the full model to low-dimensional chaos in the reduced one.
\item In the interval $5\lesssim \kappa \lesssim 8$ there is a coexistence between stable homogeneous neighborhoods (branch with low $u_{\infty}$) and a chaotic stripe at high values of $u_{\infty}$. Again, these behaviors are separated by a wide gap in the possible values of $u_{\infty}$.
\end{itemize}

\section{Conclusions} \label{conclu}

In this work nonlinear $\mathcal{B}_{\kappa}$-embeddings have been constructed that are able to yield mathematical objects with different dimensionality (scalars, vectors, matrices) and dynamical classes of models (CAs, CMLs, nonlinear PDEs, CA connections)  as a continuous parameter $\kappa \in \mathbb{R}$ is varied. We note that $\kappa$ should have a wide physical significance. If one considers, for example, many particle systems governed by statistical laws, $\kappa$ can be thought as a coarse-graining parameter, an increased value of it leading to fuzzier descriptions involving a lower number of degrees of freedom. $\kappa$ can also be considered as a scale parameter in unified field theories, involved in the connection of objects (tensors) of different dimensions at different scales in which spacetime is probed. 

When connecting two mathematical structures of different dimensions, the technique presented in this manuscript is equivalent to embedding both of them in the space in which the structure of higher dimension lives and smoothly interpolating between them in that space. The major advantage of the embeddings presented is that the embedded structures are fixed points of the embeddings, when looked at dynamically. 

Based on appropriate nonlinear $\mathcal{B}_{\kappa}$-embeddings \cite{homotopon}, a new approach to compactification in unified physical theories (e.g. supergravity in 10 or 11-dimensional spacetimes) has been suggested. The method involves no Fourier expansion and no truncation, as is usually performed on extra dimensions to account for the observable 4-dimensional universe. The limits $\kappa \to 0$ and $\kappa \to \infty$ of the $\mathcal{B}_{\kappa}$-embeddings are robust and yield the metric tensors of the spacetime with extra dimensions and the one of the observable universe, respectively.

We have also shown how $\mathcal{B}_{\kappa}$-embeddings can be used to asymptotically connect CA with nonlinear PDEs through appropriate CMLs, all these structures being particular instances of the embedding. In particular, we have shown how (nonlinear) diffusion equations naturally emerge asymptotically from this construction. This mathematical approach sheds light, therefore, on why the Laplacian operator has such a tremendous importance in physical theories, since it already emerges from the most elementary dynamical systems and interactions when the continuum limit is performed. 

In this article, we have also introduced the concept of CA connections. These are CMLs obtained from nonlinear $\mathcal{B}_{\kappa}$-embeddings, that depend on a control parameter $\kappa$ such that in the limits $\kappa \to 0$ and $\kappa \to \infty$ the CML collapses to a CA. We have shown that any two CAs in rule space can be connected in this way. A mean-field, reduced model allows a bifurcation diagram to be calculated that qualitatively captures the features observed in the spatiotemporal evolution of the connection (in those parameter regimes where the neighborhood dynamics is approximately homogeneous). We have illustrated these general results with the specific example of Wolfram elementary Boolean CA rules 30 and 110 \cite{Wolfram} constructing a connection between both rules. At intermediate $\kappa$ values, a wide variety of dynamical behavior has been observed ranging from coherent to seemingly chaotic behavior, as well as the coexistence of coherence and disorder for simple initial conditions. These behaviors have been qualitatively investigated by means of a mean-field model derived from the connection. The results presented in this article can be easily generalized to more dimensions and arbitrary order in time \cite{VGM1}.

If the parameter $\kappa$ in a CA connection is interpreted as the coupling strength on the lattice, we suggest that modulations introduced through CA connections can be used in biophysical applications to model changes in dynamical behavior induced by fuzziness or the coarsening of network connectivity \cite{VGMSMJA}. If one considers physical models of networks governed by a finite set of strict rules (CA-like), a non-vanishing value for the parameter $\kappa$ may incorporate the overall effect of the network heterogeneity, as well as the weakening of cooperative phenomena as $\kappa$ is increased. If $\kappa$ is made explicitly dependent on time, specific CA connections may also account for the effect of aging in the evolution of a system dynamics.  We, therefore, believe that the structures here introduced can be helpful to model the long time evolution of biological organisms \cite{Hwang}.

\section*{Acknowledgments} We gratefully acknowledge comments by two anonymous referees that have led to an improved version of the manuscript.

\section*{Data availability statement} 

Data sharing is not applicable to this article as no new data
were created or analysed in this study.

\end{document}